\documentstyle[amstex,manuscript,osa,cite]{revtex}

\begin{document}

%
\title{Fluctuating Elastic Rings: Statics and Dynamics}
\author{{\rm Sergey Panyukov}
\footnote[1]{Permanent address:
Theoretical department, Lebedev Physics Institute, Russian Academy
of Sciences, Moscow 117924, Russia, electronic address:
panyukov@@lpi.ac.ru}\
{\rm and Yitzhak Rabin}
\footnote[2]{Electronic address: yr@@rabinws.ph.biu.ac.il}}
\address{{\normalsize {\it Department of Physics, Bar--Ilan University}\\
\normalsize {\it Ramat-Gan 52900, Israel}}}
\maketitle
\pacs{87.15.-v, 87.15.Ya, 05.40.-a}

\begin{abstract}
We study the effects of thermal fluctuations on elastic rings.
Analytical expressions are derived for correlation functions of
Euler angles, mean square distance between points on the ring contour,
radius of gyration, and  probability distribution of writhe fluctuations.
Since fluctuation amplitudes diverge in the limit of vanishing twist
rigidity, twist elasticity is essential for the
description of fluctuating rings. We discover a
crossover from a small scale regime in which the filament behaves as a straight rod,
to a large scale regime in which spontaneous curvature is important and
twist rigidity affects the spatial configurations of the ring.
The fluctuation-dissipation relation between correlation functions of Euler angles
and response functions, is used to study the deformation of the ring by external forces.
The effects of inertia and
dissipation on the relaxation of temporal correlations of writhe
fluctuations, are analyzed using Langevin dynamics.
\end{abstract}
%

\section{Introduction}

Small circular loops of DNA (plasmids) play an important role in biological
processes such as gene transfer between bacteria and in biotechnological
applications where they are used as vectors for DNA cloning \cite{ENCYC}.
The simplest minimal model that captures both the topology and the physical
properties of such an object is that of an elastic ring that has both
bending and twist moduli. This model was used in a recent study of writhe
instability of a twisted ring\cite{White,Zajac}. However, since this work
focused on the mechanical aspects of the problem and did not consider the
effects of thermal fluctuations, it cannot be directly applied to plasmids
and other microscopic rings. The consideration of fluctuations is important
since they dominate the physics of macromolecules and determine their
statistical properties, such as characteristic dimensions, dynamics in
solution\cite{PG}, kinetics of loop formation and dissociation of short DNA
segments\cite{Volog} and molecular beacons\cite{Albert}. Recently, we
developed a theory of fluctuating elastic filaments, with arbitrary
spontaneous curvature, torsion and twist in their stress free state\cite
{PRL2000}. Since topological constraints were not taken into account in this
work, our analysis was limited to open filaments and could not be directly
applied to the study of closed objects that have the topology of a ring.

The present paper is an expanded version of a letter in which we presented
the solution of this problem for weakly fluctuating rings\cite{PRL2001}. The
analysis of reference \cite{PRL2001} is generalized to the case of
ribbonlike filaments, with two principal axes of inertia in the cross
sectional plane. We calculate the correlation functions of Euler angles, and
use them to obtain other statistical properties of fluctuating rings, such
as mean square spatial distance between points on the ring contour, and
radius of gyration. Analytical expressions for the complete probability
distribution function of writhe fluctuations and for all its moments, are
derived. A crossover length scale is found, below which straight rod
behavior dominates and the twist of the cross section with respect to the
centerline is uncorrelated with the conformation of the centerline. Above
this length scale the nonvanishing spontaneous curvature of the ring begins
to play a role and twist rigidity affects the three--dimensional
conformation of the centerline of the ring. The correlation functions of
Euler angles are used to predict the mechanical response to external torques
and forces, and to examine the effect of spontaneous orientation of the
cross section, on the deformation of ribbonlike rings. The dynamic
correlation function of writhe fluctuations is calculated in both the
inertial and the dissipative regimes. In the former case oscillatory decay
of the correlations with time is observed. When inertia is negligible, the
relaxation is monotonic and there is a transition from a short time regime
in which the relaxation rate depends only on the bending rigidity, to a long
time regime where the decay is affected by both bending and twist modes.

In section \ref{GENERAL} we present the generalized Frenet equations that
describe the conformation of a filament, and introduce the elastic energy
that governs its fluctuations about the stress free state. We express the
curvature and torsion parameters that characterize this conformation, in
terms of the Euler angles, and write down the elastic energy as a quadratic
form in the deviations of these angles from their values in the undeformed
ring. The topological constraints corresponding to a ring are introduced as
integral conditions on the fluctuations of the Euler angles, and result in
vanishing contribution of some of the lowest Fourier modes to the
fluctuation spectrum. In section \ref{SPECTRUM} we diagonalize the elastic
energy, obtain the spectrum of normal modes and discuss their physical
meaning. In section \ref{ANGULAR} we use this eigenmode expansion to
calculate the correlation functions of Euler angles. We study the dependence
of the correlators on physical parameters such as bending and twist
rigidities, and on the spontaneous orientation of the principal axes of
inertia of the cross section with respect to the plane of the ring, and
discuss the geometry of typical configurations of the ring. In section \ref
{ORIENT} we derive explicit expressions for the orientational correlation
function of the tangents to the ring, root mean square (rms) distance
between points on the ring contour, and radius of gyration, in terms of
correlation functions of Euler angles computed in the preceding sections. In
section \ref{WRITHE} we express the writhe and twist numbers that
characterize an instantaneous configuration of the ring, in terms of Euler
angles. We then use the correlation functions of Euler angles to calculate
the probability distribution function of writhe fluctuations and study its
dependence on the bending and twist rigidities. We find that the amplitude
of writhe fluctuations exhibits a crossover from small scale,
straight--rod--like regime in which twist of the cross section has no effect
on the spatial conformations of the centerline, to a large scale regime in
which the two types of fluctuations become strongly coupled due to the
spontaneous curvature of the ring. In section \ref{FDT} we use the
fluctuation--dissipation theorem that relates the previously calculated
equilibrium correlation functions of Euler angles to the response functions,
in order to study the linear response of a ring to small externally applied
forces and moments. We show that the deformation of a ribbonlike ring
depends in an essential way on the orientation of its cross section in the
undeformed reference state. In section \ref{DYNAMICS} we derive the Langevin
equations that describe both the inertial and the dissipative dynamics of
Euler angles, and use them to study the effects of bending and twist
rigidities and of the orientation of the cross section of the ribbon, on the
frequency spectrum and temporal relaxation of its writhe modes. Details of
the derivation of the Langevin equations and the calculation of the dynamic
correlation functions, are given in appendices \ref{LANGEVEN} and \ref
{SOLUTION}, respectively. In section \ref{DISCUSS} we summarize our main
results and discuss the domain of validity of our theory.

\section{General approach\label{GENERAL}}

The general theory of fluctuating noninteracting elastic filaments was
presented in Ref. \cite{PRL2000}. To each point $s$ one attaches a triad of
unit vectors $\left\{ {\bf t}_{i}(s)\right\} $ where ${\bf t}_{3}(s)$ is the
tangent vector to the curve at $s$, and the vectors ${\bf t}_{1}(s)$ and $%
{\bf t}_{2}(s)$ are directed along the axes of symmetry of the (in general,
noncircular) cross section. The spatial conformation ${\bf x}(s)$ of the
filament is given by the generalized Frenet equations 
\begin{equation}
{\displaystyle{d{\bf t}_{i} \over ds}}%
=-\sum\nolimits_{jk}e_{ijk}\omega _{j}{\bf t}_{k},  \label{Fr0}
\end{equation}
together with the inextensibility condition, 
\begin{equation}
d{\bf x}/ds={\bf t}_{3},  \label{inext}
\end{equation}
where $e_{ijk}$ is the antisymmetric unit tensor and the parameters $\left\{
\omega _{j}(s)\right\} $ characterize the curvature, torsion and twist of
the filament. The components of these vectors can be expressed in terms of
the Euler angles $\theta $, $\varphi $ and $\psi $: 
\begin{equation}
{\bf t}_{1}=\left( 
\begin{array}{c}
\cos \theta \cos \varphi \cos \psi -\sin \varphi \sin \psi \\ 
\cos \theta \sin \varphi \cos \psi +\cos \varphi \sin \psi \\ 
-\sin \theta \cos \psi
\end{array}
\right)  \label{t1}
\end{equation}
\begin{equation}
{\bf t}_{2}=\left( 
\begin{array}{c}
-\cos \theta \cos \varphi \sin \psi -\sin \varphi \cos \psi \\ 
-\cos \theta \sin \varphi \sin \psi +\cos \varphi \cos \psi \\ 
\sin \theta \sin \psi
\end{array}
\right) ,  \label{t2}
\end{equation}
\begin{equation}
{\bf t}_{3}=\left( 
\begin{array}{c}
\sin \theta \cos \varphi \\ 
\sin \theta \sin \varphi \\ 
\cos \theta
\end{array}
\right) .  \label{t3}
\end{equation}
Substituting Eqs. (\ref{t1}) -- (\ref{t3}) into Eq. (\ref{Fr0}), the Frenet
equations can be rewritten in the form: 
\begin{align}
{\displaystyle{d\theta  \over ds}}%
& =\omega _{1}\sin \psi +\omega _{2}\cos \psi ,  \nonumber \\
{\displaystyle{d\varphi  \over ds}}%
\sin \theta & =-\omega _{1}\cos \psi +\omega _{2}\sin \psi ,  \label{Frenet}
\\
{\displaystyle{d\psi  \over ds}}%
\sin \theta & =(\omega _{1}\cos \psi -\omega _{2}\sin \psi )\cos \theta
+\omega _{3}\sin \theta .  \nonumber
\end{align}
Solving these equations with respect to $\{\omega _{i}\}$ yields 
\begin{align}
\omega _{1}& =-%
{\displaystyle{d\varphi  \over ds}}%
\sin \theta \cos \psi +%
{\displaystyle{d\theta  \over ds}}%
\sin \psi ,  \label{o1} \\
\omega _{2}& =%
{\displaystyle{d\varphi  \over ds}}%
\sin \theta \sin \psi +%
{\displaystyle{d\theta  \over ds}}%
\cos \psi ,  \label{o2} \\
\omega _{3}& =%
{\displaystyle{d\psi  \over ds}}%
+\cos \theta 
{\displaystyle{d\varphi  \over ds}}%
.  \label{o3}
\end{align}

We assume that the centerline of the undeformed ring forms a circle of
radius $r$ in the $xy$ plane, and that its cross section is rotated by angle 
$\psi _{0}(s)$ around this centerline. The Euler angles that describe this
configuration are 
\begin{equation}
\theta _{0}=\pi /2,\qquad \varphi _{0}=s/r,\qquad \psi _{0}=ks/2r+\psi _{00},
\label{undef}
\end{equation}
where $k$ is an integer and $\psi _{00}$ is a constant, independent of $s$.
Eqs. (\ref{o1}) -- (\ref{o3}) can be rewritten in the form 
\begin{equation}
\omega _{01}=-\left( 1/r\right) \cos \psi _{0},\quad \omega _{02}=\left(
1/r\right) \sin \psi _{0}\text{ \quad and\quad }\omega _{03}=d\psi _{0}/ds.
\label{undef1}
\end{equation}
Although, in general, the stress free state of the ring can be arbitrarily
twisted (e.g., because of intrinsic tendency of the filament to twist), in
this work we will not consider spontaneous twist ($\omega _{03}=0$), and
taking $k=0$ we set $\psi _{0}=\psi _{00}$ (for brevity, we will denote this
constant by $\psi _{0}$ in the following). This angle characterizes the
orientation of the principal axes of the cross section with respect to the
plane of the undeformed ring. In the case of a circular cross section, all
physical observables are independent of $\psi _{0}$ and it is convenient to
set $\psi _{0}=0$.

The corresponding Euler parametrization of the triad vectors is 
\begin{equation}
{\bf t}_{01}=\left( 
\begin{array}{c}
-\sin (s/r)\sin \psi _{0} \\ 
\cos (s/r)\sin \psi _{0} \\ 
-\cos \psi _{0}
\end{array}
\right) ,\quad {\bf t}_{02}=\left( 
\begin{array}{c}
-\sin (s/r)\cos \psi _{0} \\ 
\cos (s/r)\cos \psi _{0} \\ 
\sin \psi _{0}
\end{array}
\right) ,\quad {\bf t}_{03}=\left( 
\begin{array}{c}
\cos (s/r) \\ 
\sin (s/r) \\ 
0
\end{array}
\right) .  \label{t0}
\end{equation}

In the absence of excluded--volume and other nonelastic interactions, the
energy of a filament is of purely elastic origin and can be represented as a
quadratic form in the deviations $\delta \omega _{k}=\omega _{k}-\omega
_{0k} $ \cite{White,PRL2000}, 
\begin{equation}
U=%
{\displaystyle{k_{B}T \over 2}}%
\int_{0}^{2\pi r}ds\sum\nolimits_{k=1}^{3}a_{k}\delta \omega _{k}^{2},
\label{Fdef}
\end{equation}
where $k_{B}$ is the Boltzmann constant, $T$ is the temperature, and the
bare persistence lengths $a_{k}$ represent the rigidity with respect to the
corresponding deformation modes. The above expression for the energy is
based on the linear theory of elasticity and applies to deformations whose
characteristic length scale (e.g., radius of curvature) is much larger than
the diameter of the filament\cite{Love}. Since the persistence lengths are
determined by material properties on length scales of the order of this
diameter, they are the same as those of a straight rod. We conclude that $%
a_{1}$ and $a_{2}$ are associated with the bending rigidities of the
filament with respect to the two principal axes of inertia $I_{1}$ and $%
I_{2} $ (they differ if the cross section is not circular), and that $a_{3}$
is associated with twist rigidity. In the special case of incompressible
isotropic rods with shear modulus $\mu $, the theory of elasticity yields 
\cite{Love} 
\begin{equation}
a_{1}=3\mu I_{1}/k_{B}T,\quad a_{2}=3\mu I_{2}/k_{B}T,\quad \text{and\quad }%
a_{3}=C/k_{B}T  \label{momin}
\end{equation}
where the torsional rigidity $C$ is also proportional to $\mu $ and depends
on the geometry of the cross section (for an elliptical cross section with
semi--axes $d_{1}$ and $d_{2}$, $C=\pi \mu
d_{1}^{3}d_{2}^{3}/(d_{1}^{2}+d_{2}^{2})$). In this paper we will treat $%
a_{i}$ as given material parameters of the ring.

In the following we consider only small fluctuations of the Euler angles
about their values in the undeformed state, Eq. (\ref{undef}). This
approximation remains valid as long as the bare persistence lengths are much
larger than the radius of the ring, i.e., $a_{k}\gg r$. Expanding Eqs. (\ref
{o1}) -- (\ref{o3}) in small deviations from the stress free state, we find 
\begin{align}
\delta \omega _{1}& =\left( 
{\displaystyle{\delta \psi  \over r}}%
+%
{\displaystyle{d\delta \theta  \over ds}}%
\right) \sin \psi _{0}-%
{\displaystyle{d\delta \varphi  \over ds}}%
\cos \psi _{0},  \nonumber \\
\delta \omega _{2}& =\left( 
{\displaystyle{\delta \psi  \over r}}%
+%
{\displaystyle{d\delta \theta  \over ds}}%
\right) \cos \psi _{0}+%
{\displaystyle{d\delta \varphi  \over ds}}%
\sin \psi _{0},  \label{domega} \\
\delta \omega _{3}& =%
{\displaystyle{d\delta \psi  \over ds}}%
-%
{\displaystyle{\delta \theta  \over r}}%
.  \nonumber
\end{align}

It is instructive to relate the above parameters to the curvature $\kappa $
and torsion $\tau $ familiar from differential geometry of space curves\cite
{Shape}. A circular planar ring has $\kappa _{0}=1/r$ and $\tau _{0}=0$.
Expanding in small deviations about these values yields 
\begin{equation}
\delta \kappa =%
{\displaystyle{d\delta \varphi  \over ds}}%
\quad \text{and \quad }\delta \tau =\tau =-%
{\displaystyle{\delta \theta  \over r}}%
-r%
{\displaystyle{d^{2}\delta \theta  \over ds^{2}}}%
\label{curv}
\end{equation}
As expected, fluctuations of the curvature represent bending deformations in
the plane of the ring, and depend only on the angle $\varphi $ that
describes the rotation of the tangent to the ring, in the $xy$ plane (see
Eq. (\ref{t3})). Torsion describes deviations of the filament from this
plane, and its fluctuations depend only on the deviations of the angle $%
\theta $ from $\pi /2$. The specification of the local curvature and torsion
completely determines the configuration of the centerline of any curved
filament, and the Euler angle $\psi $ complements the description by
specifying the rotation of the cross section about this centerline. However,
the elastic energy can not be factorized into a sum of contributions due to
deformation of the centerline and rotation about it. As will be shown in
Section \ref{WRITHE}, $\omega _{3}(s)$ defines the rate of twist and
therefore the persistence length $a_{3}$ is associated with twist. Twist
represents not only rotation about the centerline (the $d\psi /ds$ term in
Eq. (\ref{o3})), but also contains a contribution due to the curvature of
the centerline (the $\cos \theta d\varphi /ds$ term in the above equation).
Similarly, although inspection of Eq. (\ref{Fr0}) suggests that $\omega
_{1}(s)$ and $\omega _{2}(s)$ completely determine the variation of the
tangent ${\bf t}_{3}(s)$ as one moves along the contour, this variation
depends on the main axes of the cross section at $s$ (the vectors ${\bf t}%
_{1}(s)$ and ${\bf t}_{2}(s)$), that themselves rotate with the cross
section. This explains the $\psi $--dependence of $\omega _{1}$ and $\omega
_{2}$ in Eqs. (\ref{o1}) and (\ref{o2}). The relation between the two
descriptions ($\left\{ \theta ,\varphi ,\psi \right\} $ and $\left\{ \omega
_{i}\right\} $) is a special case of the more general relation between
Eulerian and Lagrangian descriptions in the theory of elasticity\cite
{Chaikin}. While the Euler angles describe the orientation of the triad $%
\left\{ {\bf t}_{i}(s)\right\} $ in the laboratory frame, the parameters $%
\omega _{i}(s)$ describe the local variation of this orientation as one
moves along the curve, in the frame associated with the triad itself. The
simple form of the energy, Eq. (\ref{Fdef}), is a direct consequence of this
Lagrangian description.

Substituting Eqs. (\ref{domega}) into the elastic energy, Eq. (\ref{Fdef}),
yields 
\begin{align}
U& =k_{B}T\int_{0}^{2\pi r}ds\left[ 
{\displaystyle{A_{1} \over 2}}%
\left( 
{\displaystyle{d\delta \theta  \over ds}}%
+%
{\displaystyle{\delta \psi  \over r}}%
\right) ^{2}+%
{\displaystyle{A_{2} \over 2}}%
\left( 
{\displaystyle{d\delta \varphi  \over ds}}%
\right) ^{2}\right.  \nonumber \\
& \left. +A_{3}\left( 
{\displaystyle{d\delta \theta  \over ds}}%
+%
{\displaystyle{\delta \psi  \over r}}%
\right) 
{\displaystyle{d\delta \varphi  \over ds}}%
+%
{\displaystyle{a_{3} \over 2}}%
\left( 
{\displaystyle{d\delta \psi  \over ds}}%
-%
{\displaystyle{\delta \theta  \over r}}%
\right) ^{2}\right] ,  \label{Fr}
\end{align}
where the coefficients $A_{i}$ are defined as, 
\begin{eqnarray}
A_{1} &=&a_{1}\cos ^{2}\psi _{0}+a_{2}\sin ^{2}\psi _{0},\qquad
A_{2}=a_{1}\sin ^{2}\psi _{0}+a_{2}\cos ^{2}\psi _{0},  \nonumber \\
A_{3} &=&(a_{2}-a_{1})\cos \psi _{0}\sin \psi _{0}.  \label{A123}
\end{eqnarray}
For $a_{2}>a_{1}$, the constant Euler angle $\psi _{0}$ measures the angle
between the major axis of inertia and the $xy$ plane. The case $\psi _{0}=0$
($\psi _{0}=\pi /2)$ corresponds to major axis that lies in the $xy$ plane
(normal to the $xy$ plane). The coefficients $A_{i}$ obey the relations 
\begin{equation}
A_{1}A_{2}-A_{3}^{2}=a_{1}a_{2},\qquad A_{1}+A_{2}=a_{1}+a_{2}  \label{Aa}
\end{equation}

The periodic boundary conditions on the Euler angles 
\begin{equation}
\delta \theta (2\pi r)=\delta \theta (0),\qquad \delta \psi \left( 2\pi
r\right) =\delta \psi (0),\qquad \delta \varphi (2\pi r)=\delta \varphi (0)
\label{bound0}
\end{equation}
are supplemented by the condition that the ring is closed in three
dimensional space, ${\bf x}(2\pi r)={\bf x}(0)$. Using Eq. (\ref{inext})
this condition can be recast into an integral form, 
\begin{equation}
\int_{0}^{2\pi r}ds\delta {\bf t}_{3}(s)=0.  \label{close}
\end{equation}

For small deviations from equilibrium we get from Eq. (\ref{t3}), 
\begin{equation}
\delta {\bf t}_{3}(s)=\left( 
\begin{array}{c}
-\delta \varphi (s)\sin (s/r) \\ 
\delta \varphi (s)\cos (s/r) \\ 
-\delta \theta (s)
\end{array}
\right) ,  \label{dt3}
\end{equation}
and the boundary conditions can be written as 
\begin{equation}
\int_{0}^{2\pi r}ds\delta \theta (s)=\int_{0}^{2\pi r}ds\delta \varphi
(s)\cos (s/r)=\int_{0}^{2\pi r}ds\delta \varphi (s)\sin (s/r)=0.
\label{bound}
\end{equation}
Since the deviations of the Euler angles are periodic functions of $s$, they
can be expanded in Fourier series 
\begin{equation}
\delta \eta (s)=\sum\nolimits_{n}\tilde{\eta}(n)e^{ins/r},\qquad \tilde{\eta}%
(-n)=\tilde{\eta}^{\ast }(n),\qquad  \label{exp}
\end{equation}
for each $\eta =\theta ,\varphi ,\psi $, where the sum goes over all
positive and negative integers $n$. The boundary conditions, Eqs. (\ref
{bound}), can be expressed as conditions on the Fourier coefficients, 
\begin{equation}
\tilde{\theta}(0)=\tilde{\varphi}(1)=0.  \label{bound1}
\end{equation}

Substituting Eqs. (\ref{exp}) into Eq. (\ref{Fr}) we find, 
\begin{align}
{\displaystyle{U \over 2\pi rk_{B}T}}%
& =%
{\displaystyle{1 \over r^{2}}}%
\left[ 
{\displaystyle{A_{1} \over 2}}%
\left| \tilde{\psi}(0)\right| ^{2}+\left( A_{1}+a_{3}\right) \left| i\tilde{%
\theta}(1)+\tilde{\psi}(1)\right| ^{2}\right] +  \label{FreeEn} \\
& 
{\displaystyle{1 \over r^{2}}}%
\sum_{n=2}^{\infty }\left\{ A_{1}\left| in\tilde{\theta}(n)+\tilde{\psi}%
(n)\right| ^{2}+A_{2}n^{2}\left| \tilde{\varphi}(n)\right| ^{2}\right. 
\nonumber \\
& \left. -2A_{3}\left[ in\tilde{\theta}(n)+\tilde{\psi}(n)\right] in\tilde{%
\varphi}(-n)+a_{3}\left| in\tilde{\psi}(n)-\tilde{\theta}(n)\right|
^{2}\right\} .  \nonumber
\end{align}
The energy does not depend on modes $\tilde{\psi}(1)=-i\tilde{\theta}(1)$
and $\tilde{\varphi}(0)$ that correspond to rigid body rotation of the
entire ring, with respect to axes lying in the plane of the ring and normal
to it, respectively.

The quadratic form inside the sum over $n$ in Eq. (\ref{FreeEn}) can be
represented as a matrix in the space spanned by the Fourier components $%
\tilde{\theta}(n),$ $\tilde{\varphi}(n)$ and $\tilde{\psi}(n)$ (this applies
to $n>1;$ the cases $n=0,\pm 1$ will be considered separately), 
\begin{equation}
{\bf Q}(n)=\left( 
\begin{array}{ccc}
A_{1}n^{2}+a_{3} & A_{3}n^{2} & -i\left( A_{1}+a_{3}\right) n \\ 
A_{3}n^{2} & A_{2}n^{2} & -iA_{3}n \\ 
i\left( A_{1}+a_{3}\right) n & iA_{3}n & a_{3}n^{2}+A_{1}
\end{array}
\right) .  \label{F1}
\end{equation}

\section{Spectrum of fluctuations\label{SPECTRUM}}

In order to obtain the spectrum of fluctuations of the ring, we diagonalize
the free energy, Eq. (\ref{FreeEn}), by expanding the Fourier components $%
\tilde{\eta}(n)$ in the eigenvectors $\eta _{k}(n)$ of the matrix ${\bf Q}%
(n) $, 
\begin{equation}
\tilde{\eta}(n)=\sum\nolimits_{k}c_{k}(n)\eta _{k}(n),  \label{modes}
\end{equation}
where $\eta =\theta ,\varphi ,\psi $ and $\eta _{k}(n)$ is the $\eta $--th
component of the eigenvector ${\bf \eta }_{k}(n)=\{\theta _{k}(n),\varphi
_{k}(n),\psi _{k}(n)\}$ of the quadratic form, Eq. (\ref{FreeEn}),
corresponding to the eigenvalue $\lambda _{k}(n)$. They are normalized by
the conditions, 
\begin{equation}
\sum\nolimits_{\eta }\eta _{k}(n)\eta _{l}\left( -n\right) =\delta
_{kl},\qquad \sum\nolimits_{k}\eta _{k}(n)\eta _{k}^{^{\prime }}\left(
-n\right) =\delta _{\eta \eta ^{\prime }}.  \label{norm}
\end{equation}
Expanding the elastic energy in the eigenmodes gives 
\begin{equation}
U=%
{\displaystyle{\pi k_{B}T \over r}}%
\sum_{n=0}^{\infty }\sum_{k}\lambda _{k}(n)\left| c_{k}(n)\right| ^{2}.
\label{Fl}
\end{equation}
The three eigenvalues $\lambda _{k}(n)$ corresponding to the Fourier mode $n$%
, are the roots of the characteristic cubic polynomial, 
\begin{equation}
\lambda ^{3}-b_{2}\lambda ^{2}+b_{1}\lambda -b_{0}=0,  \label{secular}
\end{equation}
with coefficients 
\begin{align}
b_{0}& =a_{1}a_{2}a_{3}n^{2}\left( n^{2}-1\right) ^{2},  \nonumber \\
b_{1}& =\left( a_{1}a_{2}+a_{2}a_{3}+a_{1}a_{3}\right) n^{2}\left(
n^{2}+1\right) -A_{1}a_{3}\left( 3n^{2}-1\right) ,  \label{ci} \\
b_{2}& =\left( a_{1}+a_{2}+a_{3}\right) n^{2}+A_{1}+a_{3},  \nonumber
\end{align}
where we used Eqs. (\ref{A123}) and (\ref{Aa}) to simplify cumbersome
mathematical expressions. Since the matrix ${\bf Q}(n)$\ is Hermitian, its
eigenvalues are real.

Inspection of Eqs. (\ref{secular}) and (\ref{ci}) shows that $\lambda
_{k}(-n)=\lambda _{k}(n)$ and that all eigenvalues with $n>1$ are positive.
Because of the boundary conditions, Eqs. (\ref{bound1}), there are only two
independent normal modes corresponding to each of the cases, $n=0$ and $n=1$%
. In order to understand the physical meaning of these modes, we introduce
the components of the Fourier transforms of the curvature and torsion, Eq. (%
\ref{curv}), 
\begin{equation}
\tilde{\kappa}(n)=%
{\displaystyle{in \over r}}%
\tilde{\varphi}(n)\quad \text{and \quad }\tilde{\tau}(n)=%
{\displaystyle{n^{2}-1 \over r}}%
\tilde{\theta}(n).  \label{curv1}
\end{equation}
Substituting the boundary conditions $\tilde{\theta}(0)=\tilde{\varphi}(1)=0$
into the above expressions we conclude that for modes with $n=0$ and $1$,
both $\delta \kappa $ and $\delta \tau $ vanish and, therefore, these modes
do not affect the planar circular configuration of the centerline of the
ring. There are two zero energy modes that correspond to symmetry operations
on the undeformed ring. One $n=0$ mode, with eigenfunction $\varphi
_{1}(0)=1 $; $\psi _{1}(0)=0$, describes the rotation of the ring about the $%
z$ axis. One $n=1$ mode, with eigenfunction $\theta _{1}(1)=1$; $\psi
_{1}(1)=-i$, corresponds to rotation of the ring about an axis in the $xy$
plane. The two remaining modes have an energy gap and are twist modes that
leave the centerline undisturbed. The $n=0$ mode with eigenfunction $\varphi
_{2}(0)=0$; $\psi _{2}(0)=1$ has an eigenvalue $\lambda _{2}(0)=A_{1}$, and
describes uniform twist of the ring about its centerline. Since this
eigenvalue does not vanish for arbitrary $a_{1}\ $and $a_{2}$, we conclude
that uniform twist of a ring costs energy{\em \ }even if the ring has a
circular cross section. This conclusion agrees with reference \cite{Tobias1}%
, where the dynamics of the uniform twist mode was studied. The $n=1$ mode
with eigenfunction $\theta _{2}(1)=1$; $\psi _{2}(1)=i$ has the eigenvalue $%
\lambda _{2}(1)=2(A_{1}+a_{3})$ and corresponds to rotation of the ring with
respect to an axis that passes through the centerline and lies in the $xy$
plane, accompanied by twist of the cross section by the angle $\psi $ that
varies periodically (as $\cos (s/r)$) along the contour of the ring. The
dynamics of this mode was studied in reference \cite{Tobias2}.

In the limit $n\gg 1$, fluctuations of the three Euler angles are decoupled
and $\lambda _{k}(n)\simeq a_{k}n^{2}$. In general, each normal mode of the
ring corresponds to fluctuations of all three Euler angles, $\delta \theta
(s)$, $\delta \varphi (s)$ and $\delta \psi (s)$, and describes a complex
three--dimensional configuration.

The eigenvalue problem is simplified for a circular cross section ($%
a_{2}=a_{1}$), or when the cross section is asymmetric but $\psi _{0}=0$
(the case $\psi _{0}=\pi /2$ is reduced to $\psi _{0}=0$ by the substitution 
$a_{1}\leftrightarrow a_{2}$). In these cases the mode $\delta \varphi (s)$
decouples from the other two modes and has the spectrum $\lambda
_{1}(n)=a_{1}n^{2}$ ($n\neq \pm 1$). This mode corresponds to bending
fluctuations that lie entirely in the plane of the ring. The other two modes
are linear combinations of $\delta \theta (s)$ and $\delta \psi (s)$, with
eigenvalues 
\begin{equation}
\lambda _{2,3}(n)=%
{\displaystyle{a_{2}+a_{3} \over 2}}%
(n^{2}+1)\pm \sqrt{\left( 
{\displaystyle{a_{2}-a_{3} \over 2}}%
\right) ^{2}(n^{2}+1)^{2}+4n^{2}a_{2}a_{3}}.  \label{quadr}
\end{equation}
Eq. (\ref{quadr}) can be further simplified in the limit of large rigidity
with respect to twist, $a_{3}\gg a_{2}$, in which case 
\begin{equation}
\begin{array}{ccc}
\lambda _{2}(n)=a_{2}%
{\displaystyle{(n^{2}-1)^{2} \over n^{2}+1}}%
& \text{for} & n\geq 1, \\ 
\lambda _{3}(n)=a_{3}(n^{2}+1) & \text{for} & n\geq 2.
\end{array}
\label{l23}
\end{equation}
In the opposite limit $a_{3}\ll a_{2}$, the eigenvalues can be found by
substituting $a_{2}\leftrightarrow a_{3}$ in Eq. (\ref{l23}).

Inspection of Eq. (\ref{quadr}) shows that $\lambda _{3}(n)$\ vanishes
identically for all $n$\ when $a_{3}=0$\ (this statement applies even to
rings with noncircular cross section -- see Eqs. (\ref{secular}) and (\ref
{ci})), indicating that the amplitudes of the corresponding fluctuation
modes grow without limit in the absence of twist rigidity. Examining the
expression for the elastic energy, Eq. (\ref{Fr}), we conclude that these
zero energy modes correspond to fluctuations for which 
\begin{equation}
d\delta \theta /ds=-\delta \psi /r.  \label{zeroE}
\end{equation}
\ In the absence of twist rigidity, twist fluctuations carry no energy
penalty and the angle of twist of the cross section ($\delta \psi )$ can
always adjust itself to arbitrary deviation of the centerline from the plane
of the unperturbed ring ($\delta \theta )$, so that this condition, Eq. (\ref
{zeroE}), its satisfied. The presence of an infinite number of zero energy
modes means that twist rigidity ($a_{3}\neq 0$) is absolutely essential for
stabilizing the ring against out--of--plane fluctuations, and that bending
elasticity alone can not suppress this instability.

\section{Correlations of Euler angles\label{ANGULAR}}

Applying the equipartition theorem to Eq. (\ref{Fl}), we get 
\begin{equation}
\left\langle c_{k}(n)c_{k^{\prime }}(-n^{\prime })\right\rangle =%
{\displaystyle{r \over \pi \lambda _{k}(n)}}%
\delta _{nn^{\prime }}\delta _{kk^{\prime }}.  \label{cav}
\end{equation}
Using expansion (\ref{modes}) and averaging with the help of Eq. (\ref{cav}%
), the correlation functions of Euler angles can be expressed in terms of
the eigenvalues $\lambda _{k}(n)$ and the eigenfunctions $\eta _{k}(n)$ of
the ${\bf Q}(n)$ matrix: 
\begin{equation}
\left\langle \delta \eta \left( s\right) \delta \eta ^{\prime }\left(
s^{\prime }\right) \right\rangle =\sum_{n}e^{in\left( s-s\prime \right)
/r}\left\langle \tilde{\eta}(n)\tilde{\eta}^{\prime }(-n)\right\rangle =%
{\displaystyle{r \over \pi }}%
\sum_{n}e^{in\left( s-s\prime \right) /r}\sum_{k}%
{\displaystyle{\eta _{k}(n)\eta _{k}^{\prime }(-n) \over \lambda _{k}(n)}}%
,  \label{ff}
\end{equation}
where $\delta \eta ,\delta \eta ^{\prime }=\delta \theta $, $\delta \varphi $%
, $\delta \psi $. Care should be exercised in evaluating the above
expression, when considering the contribution of the modes with $n=0,\pm 1$,
since modes with vanishing eigenvalues should be excluded. A straightforward
calculation gives 
\begin{align}
& \sum_{n=0,\pm 1}e^{in\left( s-s^{\prime }\right) /r}\sum_{k}%
{\displaystyle{{\bf \eta }_{k}(n){\bf \eta }_{k}(-n) \over \lambda _{k}(n)}}%
\nonumber \\
& =%
{\displaystyle{1 \over A_{1}}}%
\left( 
\begin{array}{ccc}
0 & 0 & 0 \\ 
0 & 0 & 0 \\ 
0 & 0 & 1
\end{array}
\right) +%
{\displaystyle{1 \over A_{1}+a_{3}}}%
\left( 
\begin{array}{ccc}
\cos \left( 
{\displaystyle{s-s^{\prime } \over r}}%
\right) & 0 & -\sin \left( 
{\displaystyle{s-s^{\prime } \over r}}%
\right) \\ 
0 & 0 & 0 \\ 
\sin \left( 
{\displaystyle{s-s^{\prime } \over r}}%
\right) & 0 & \cos \left( 
{\displaystyle{s-s^{\prime } \over r}}%
\right)
\end{array}
\right) ,  \label{n01}
\end{align}
where ${\bf \eta }_{k}{\bf \eta }_{k}$ denotes the direct product of two
vectors ${\bf \eta }_{k}$, the $\eta \eta ^{\prime }$ component of which is $%
\eta _{k}\eta _{k}^{\prime }$.

For $n\neq 0,\pm 1$ we find 
\begin{equation}
\sum_{k}%
{\displaystyle{\eta _{k}(n)\eta _{k}^{\prime }\left( -n\right)  \over \lambda _{k}(n)}}%
=Q_{\eta \eta ^{\prime }}^{-1}(n),  \label{n23}
\end{equation}
where ${\bf Q}^{-1}(n)$ is the inverse of the matrix ${\bf Q}$ defined in
Eq. (\ref{F1}), 
\begin{equation}
{\bf Q}^{-1}(n)=\left( 
\begin{array}{ccc}
\frac{1}{a_{3}(n^{2}-1)^{2}}+\frac{n^{2}}{a_{\bot }(n^{2}-1)^{2}} & \frac{%
-A_{3}}{a_{1}a_{2}(n^{2}-1)} & \left( \frac{1}{a_{3}}+\frac{1}{a_{\bot }}%
\right) \frac{in}{(n^{2}-1)^{2}} \\ 
\frac{-A_{3}}{a_{1}a_{2}(n^{2}-1)} & \frac{1}{a_{\Vert }n^{2}} & \frac{%
-iA_{3}}{a_{1}a_{2}(n^{2}-1)} \\ 
-\left( \frac{1}{a_{3}}+\frac{1}{a_{\bot }}\right) \frac{in}{(n^{2}-1)^{2}}
& \frac{iA_{3}}{a_{1}a_{2}n(n^{2}-1)} & \frac{n^{2}}{a_{3}(n^{2}-1)^{2}}+%
\frac{1}{a_{\bot }(n^{2}-1)^{2}}
\end{array}
\right) ,  \label{M2}
\end{equation}
and where 
\begin{equation}
{\displaystyle{1 \over a_{\bot }}}%
=%
{\displaystyle{\cos ^{2}\psi _{0} \over a_{1}}}%
+%
{\displaystyle{\sin ^{2}\psi _{0} \over a_{2}}}%
,\qquad 
{\displaystyle{1 \over a_{\Vert }}}%
=%
{\displaystyle{\sin ^{2}\psi _{0} \over a_{1}}}%
+%
{\displaystyle{\cos ^{2}\psi _{0} \over a_{2}}}%
.  \label{ainv}
\end{equation}
Effective persistence lengths $a_{3}$ and $a_{\bot }$ control both
fluctuations perpendicular to the plane of the ring and fluctuations of the
twist angle $\psi $, and $a_{\Vert }$ controls fluctuations in the plane of
the ring. Using Eqs. (\ref{ff}) and (\ref{M2}), we obtain the correlation
functions of Euler angles (here $s=\left| s_{2}-s_{1}\right| ,\quad 0<s<2\pi
r$) {\bf \ } 
\begin{align}
\left\langle \delta \theta (s_{1})\delta \theta \left( s_{2}\right)
\right\rangle & =%
{\displaystyle{r \over \pi }}%
{\displaystyle{\cos (s/r) \over A_{1}+a_{3}}}%
+%
{\displaystyle{r \over \pi a_{3}}}%
f_{3}\left( 
{\displaystyle{s \over r}}%
\right) +%
{\displaystyle{r \over \pi a_{\bot }}}%
f_{1}\left( 
{\displaystyle{s \over r}}%
\right) ,  \nonumber \\
\left\langle \delta \varphi (s_{1})\delta \varphi \left( s_{2}\right)
\right\rangle & =%
{\displaystyle{r \over \pi a_{\Vert }}}%
f_{2}\left( 
{\displaystyle{s \over r}}%
\right) ,  \nonumber \\
\left\langle \delta \psi (s_{1})\delta \psi (s_{2})\right\rangle & =%
{\displaystyle{r \over \pi A_{1}}}%
+%
{\displaystyle{r \over \pi }}%
{\displaystyle{\cos \left( s/r\right)  \over A_{1}+a_{3}}}%
+%
{\displaystyle{r \over \pi a_{3}}}%
f_{1}\left( 
{\displaystyle{s \over r}}%
\right) +%
{\displaystyle{r \over \pi a_{\bot }}}%
f_{3}\left( 
{\displaystyle{s \over r}}%
\right) ,  \label{ang} \\
\left\langle \delta \theta (s_{1})\delta \varphi \left( s_{2}\right)
\right\rangle & =-%
{\displaystyle{r\sin \left( 2\psi _{0}\right)  \over 2\pi }}%
\left( 
{\displaystyle{1 \over a_{2}}}%
-%
{\displaystyle{1 \over a_{1}}}%
\right) \left[ f_{1}\left( 
{\displaystyle{s \over r}}%
\right) -f_{3}\left( 
{\displaystyle{s \over r}}%
\right) \right] ,  \nonumber \\
\left\langle \delta \theta (s_{1})\delta \psi (s_{2})\right\rangle & =-%
{\displaystyle{r \over \pi }}%
{\displaystyle{\sin (s/r) \over A_{1}+a_{3}}}%
+%
{\displaystyle{r \over \pi }}%
\left( \frac{1}{a_{3}}+\frac{1}{a_{\bot }}\right) f_{4}\left( 
{\displaystyle{s \over r}}%
\right) ,  \nonumber \\
\left\langle \delta \varphi (s_{1})\delta \psi \left( s_{2}\right)
\right\rangle & =-%
{\displaystyle{r\sin \left( 2\psi _{0}\right)  \over 2\pi }}%
\left( 
{\displaystyle{1 \over a_{2}}}%
-%
{\displaystyle{1 \over a_{1}}}%
\right) f_{5}\left( 
{\displaystyle{s \over r}}%
\right) ,  \nonumber
\end{align}
where we defined, for $0<x<2\pi ,$%
\begin{align}
f_{1}(x)& =\sum_{n=2}^{\infty }%
{\displaystyle{n^{2}\cos nx \over \left( n^{2}-1\right) ^{2}}}%
=\left[ 
{\displaystyle{\left( \pi -x\right) ^{2} \over 8}}%
-%
{\displaystyle{\pi ^{2} \over 24}}%
+%
{\displaystyle{1 \over 16}}%
\right] \cos x-%
{\displaystyle{\pi -x \over 4}}%
\sin x,  \nonumber \\
f_{2}(x)& =\sum_{n=2}^{\infty }%
{\displaystyle{\cos nx \over n^{2}}}%
=%
{\displaystyle{\left( \pi -x\right) ^{2} \over 4}}%
-\frac{\pi ^{2}}{12}-\cos x,  \nonumber \\
f_{3}(x)& =\sum_{n=2}^{\infty }%
{\displaystyle{\cos nx \over \left( n^{2}-1\right) ^{2}}}%
=\left[ 
{\displaystyle{\left( \pi -x\right) ^{2} \over 8}}%
-%
{\displaystyle{\pi ^{2} \over 24}}%
-%
{\displaystyle{3 \over 16}}%
\right] \cos x+%
{\displaystyle{\pi -x \over 4}}%
\sin x-%
{\displaystyle{1 \over 2}}%
,  \label{fi} \\
f_{4}(x)& =\sum_{n=2}^{\infty }\frac{n\sin nx}{\left( n^{2}-1\right) ^{2}}=%
\left[ 
{\displaystyle{\left( \pi -x\right) ^{2} \over 8}}%
-%
{\displaystyle{\pi ^{2} \over 24}}%
+%
{\displaystyle{1 \over 16}}%
\right] \sin x,  \nonumber \\
f_{5}(x)& =\sum_{n=2}^{\infty }%
{\displaystyle{\sin nx \over n\left( n^{2}-1\right) }}%
=%
{\displaystyle{3 \over 4}}%
\sin x+%
{\displaystyle{\pi -x \over 2}}%
\left( \cos x-1\right) .  \nonumber
\end{align}

Inspection of Eqs. (\ref{ang}) shows that the bare persistence length
associated with the twist rigidity, $a_{3}$, plays a fundamentally important
role: fluctuations of the angles $\psi $ and $\theta $ and the correlation
between these angles, diverge in the limit $a_{3}\rightarrow 0$! Therefore,
simplified models of elastic filaments with nonvanishing spontaneous
curvature that do not take into account twist rigidity, can not describe
fluctuations and elastic response of the ring. This is not the case for a
straight rod, whose spatial fluctuations can be successfully described by
the wormlike chain model\cite{Mezard} (with $a_{3}=0$). The reason for the
difference stems from the fact that the elastic energy of straight rods
contains no coupling between the angles that describe the spatial
conformation of the centerline ($\theta $\ and $\varphi $) and the angle
that describes the twist of the cross section about this centerline ($\psi
). $\ When twist rigidity vanishes ($a_{3}=0$) there is no energy penalty
for twisting the cross section about the centerline and the amplitude of
twist fluctuations of the cross section about the centerline diverges, but
the presence of bending rigidity ($a_{1},a_{2}\neq 0$) suffices to suppress
spatial fluctuations of the centerline about its straight stress free
configuration. For rings, the elastic energy in Eq. (\ref{Fr}) contains
cross terms in the angles $\delta \psi $\ and $\delta \theta $\ that couple
both types of fluctuations. Inspection of Eq. (\ref{Fr}) shows that when $%
a_{3}=0$,\ fluctuations with $d\delta \theta /ds+\delta \psi /r=0$\ have
zero energy cost (see Eq. (\ref{zeroE})) and, since in the absence of twist
rigidity the angle $\delta \psi $ can always adjust itself to satisfy the
condition $\delta \psi =-r$ $d\delta \theta /ds$, for $a_{3}=0$\ there is no
elastic energy penalty for out--of--plane fluctuations of the ring and the
amplitude of such fluctuations diverges. We conclude that standard wormlike
chain theories in which only bending rigidity is taken into account, can not
model fluctuating rings.

In Figs. 1 -- 2 we plot correlation functions of Euler angles, for a ring
with circularly symmetric cross section. Substituting $a_{1}=a_{2}$ in the
expressions for the angular correlators in Eqs. (\ref{ang}) we find $%
\left\langle \delta \theta (s_{1})\delta \varphi \left( s_{2}\right)
\right\rangle =\left\langle \delta \varphi (s_{1})\delta \psi \left(
s_{2}\right) \right\rangle =0$. The physical reason for this behavior
becomes clear when one recalls the discussion of the eigenvalue problem for
a ring with circularly symmetric cross section (see Eq. (\ref{quadr})). In
this case, fluctuations of $\varphi (s)$ decouple from those of the other
two angles and therefore, cross correlation functions involving $\delta
\varphi $ vanish identically. In Fig. 1 we consider the case $a_{1}=a_{2}\ll
a_{3}$, i.e., twist rigidity is much larger than the that of the bending
modes. The diagonal angular correlation functions are oscillatory functions
of the contour distance, with maxima at $\left| s_{2}-s_{1}\right| =0,$ $\pi
r$ and $2\pi r$ (they are symmetric with respect to reflection about the
point $\left| s_{2}-s_{1}\right| =\pi r$). These behaviors result from
interference of two wave packets propagating along two opposite directions
along the ring. As a consequence of the large twist rigidity, the correlator
of the twist angle is always positive, while $\left\langle \delta \theta
(s_{1})\delta \theta \left( s_{2}\right) \right\rangle $ and $\left\langle
\delta \varphi (s_{1})\delta \varphi \left( s_{2}\right) \right\rangle $
fluctuate around zero. The cross correlation function, $\left\langle \delta
\theta (s_{1})\delta \psi (s_{2})\right\rangle $, vanishes as $\left|
s_{2}-s_{1}\right| \rightarrow 0$. The physical reason for this surprising
behavior is that a short segment of the ring confined between these points
can be considered as a nearly straight incompressible rod. Since twist of
such a rod does not produce any deformation, local fluctuations of twist and
of the other two modes are not correlated with each other. For larger
contour separations, spontaneous curvature begins to play a role and
fluctuations of $\theta $ and $\psi $ become coupled$.$ This is a
manifestation of the crossover from small scale (twist and spatial
conformation fluctuate independently) to large scale (coupled twist and
centerline fluctuations) behavior, that will be discussed in greater detail
in section \ref{WRITHE}.

In Fig. 2 we present the case of small twist rigidity, $a_{1}=a_{2}\gg a_{3}$%
. The twist correlation function develops four nodes (i.e., points at which
it vanishes) and, at the same time, its amplitude is strongly enhanced. In
Fig. 2 we did not plot the correlation function $\left\langle \delta \varphi
(s_{1})\delta \varphi \left( s_{2}\right) \right\rangle $, since it depends
only on the bending rigidities (see the second of Eqs. (\ref{ang})) and is
therefore the same as in Fig. 1. Fig. 3 deals with the case of an asymmetric
cross section (or asymmetric rigidity in the cross sectional plane), $%
a_{1}\neq a_{2}$. The cross correlations $\left\langle \delta \theta
(s_{1})\delta \varphi \left( s_{2}\right) \right\rangle \ $and $\left\langle
\delta \varphi (s_{1})\delta \psi \left( s_{2}\right) \right\rangle $ no
longer vanish (for $\psi _{0}\neq 0,\pi /2$), even though their amplitude is
much smaller than that of $\left\langle \delta \theta (s)\delta \psi
(0)\right\rangle $. Since the arguments presented in the preceding paragraph
apply here as well, the two cross correlation functions involving $\delta
\psi $ vanish as $s_{2}\rightarrow s_{1}$.$\ $The cross correlation function 
$\left\langle \delta \theta (s_{1})\delta \varphi \left( s_{2}\right)
\right\rangle $ behaves in a way similar to that of the diagonal correlation
functions and is symmetric about $\left| s_{2}-s_{1}\right| =\pi r$.

We would like to comment on the physical meaning of fluctuations of the
angle $\varphi (s)$. We find from Eq. (\ref{ang}) 
\begin{equation}
\left\langle \left[ \delta \varphi (s_{2})-\delta \varphi (s_{1})\right]
^{2}\right\rangle =%
{\displaystyle{s_{\parallel } \over a_{\Vert }}}%
-%
{\displaystyle{2r \over \pi }}%
{\displaystyle{1 \over a_{\Vert }}}%
\left[ 1-2\cos \left( 
{\displaystyle{s \over r}}%
\right) \right] ,  \label{phin}
\end{equation}
where the ``parallel'' persistence length $a_{\Vert }$ is defined in Eq. (%
\ref{ainv}), and where $s_{\parallel }=s(1-s/2\pi r)$ is the effective
contour length for parallel connection of two segments, one of length $%
s=\left| s_{2}-s_{1}\right| $ and the second of length $2\pi r-s$
(analogously to parallel connection of resistors in an electrical circuit).
The effective elastic modulus between points $s_{1}$ and $s_{2}$ is
proportional to 
\begin{equation}
\frac{1}{s_{\parallel }}=\frac{1}{s}+\frac{1}{2\pi r-s},\qquad \text{or}%
\qquad s_{\parallel }=s\left( 1-\frac{s}{2\pi r}\right) .  \label{spar}
\end{equation}
The second term on the rhs of Eq. (\ref{phin}) arises due to subtraction of
the contribution of the mode $\tilde{\varphi}(1)$ because of the closure of
the ring. Eq. (\ref{phin}) describes the Brownian fluctuations of phase $%
\varphi (s)$ on a circle, with effective ``diffusion'' coefficient $a_{\Vert
}^{-1}$. This means that the angle $\varphi $ can jump discontinuously from
point to point and therefore, the amplitude of its derivative $d\varphi /ds$
diverges. Since $d\varphi /ds$ is the local curvature of the filament (see
Eq. (\ref{curv})), we conclude that $\left\langle \left[ \delta \kappa (s)%
\right] ^{2}\right\rangle $ $\rightarrow \infty $. A similar calculation for
the second derivative of the angle $\theta $ shows that its amplitude
diverges and therefore $\left\langle \left[ \delta \tau (s)\right]
^{2}\right\rangle $ $\rightarrow \infty $ as well. The above divergences are
eliminated by a cutoff on length scales of the order of the thickness of the
filament and, on length scales larger than this diameter, the contour of the
ring remains a smooth and continuous curve in the process of thermal
fluctuations.

\section{Spatial correlations and radius of gyration\label{ORIENT}}

We proceed to calculate the correlation function $\left\langle \left[ {\bf x}%
(s_{1})-{\bf x}\left( s_{2}\right) \right] ^{2}\right\rangle $ that measures
the mean square spatial separation between points $s_{1}$ and $s_{2}$ on the
contour of the filament. Integrating Eq. (\ref{inext}), yields ${\bf x}%
(s_{1})-{\bf x}(s_{2})=\int_{s_{2}}^{s_{1}}{\bf t}_{3}(s^{\prime
})ds^{\prime }$ and we can express this correlation function in terms of the
correlator of tangents to the ring, at two arbitrary points on the contour, $%
\left\langle {\bf t}_{3}(s^{\prime })\bullet {\bf t}_{3}(s^{\prime \prime
})\right\rangle $. We show below that this orientational correlation
function of the tangent vectors can be expressed in terms of correlation
functions of Euler angles. Expanding the vector ${\bf t}_{3}$ to second
order in deviations of Euler angles, $\delta \eta $, from their unperturbed
values gives, 
\begin{equation}
{\bf t}_{3}\equiv {\bf t}_{03}+\delta {\bf t}_{3}=\delta \theta {\bf t}%
_{01}^{\prime }{\bf +}\delta \varphi {\bf t}_{02}^{\prime }+\left[ 1-%
{\displaystyle{1 \over 2}}%
\left( \delta \theta ^{2}+\delta \varphi ^{2}\right) \right] {\bf t}_{03},
\label{t30}
\end{equation}
where the vectors ${\bf t}^{\prime}_{0i}(s)$ are defined by 
\begin{equation}
{\bf t}_{01}^{\prime }(s)=\left( 
\begin{array}{c}
0 \\ 
0 \\ 
-1
\end{array}
\right) ,\qquad {\bf t}_{02}^{\prime }(s)=\left( 
\begin{array}{c}
-\sin (s/r) \\ 
\cos (s/r) \\ 
0
\end{array}
\right) ,\qquad {\bf t}_{03}(s)=\left( 
\begin{array}{c}
\cos (s/r) \\ 
\sin (s/r) \\ 
0
\end{array}
\right) .  \label{t123}
\end{equation}
When $\psi _{0}=0$, these vectors coincide with the vectors of unperturbed
triad, Eq. (\ref{t0}). Using Eq. (\ref{t30}) we find (in matrix notation) 
\begin{align}
\left\langle {\bf t}_{3}(s_{1}){\bf t}_{3}(s_{2})\right\rangle & =\left(
1-\left\langle \delta \theta ^{2}\right\rangle -\left\langle \delta \varphi
^{2}\right\rangle \right) {\bf t}_{03}(s_{1}){\bf t}_{03}(s_{2})  \nonumber
\\
& +\left\langle \delta \theta (s_{1})\delta \theta (s_{2})\right\rangle {\bf %
t}_{01}^{\prime }(s_{1}){\bf t}_{01}^{\prime }(s_{2})+\left\langle \delta
\varphi (s_{1})\delta \varphi (s_{2})\right\rangle {\bf t}_{02}^{\prime
}(s_{1}){\bf t}_{02}^{\prime }(s_{2})  \label{t33} \\
& +\left\langle \delta \theta (s_{1})\delta \varphi (s_{2})\right\rangle 
{\bf t}_{01}^{\prime }(s_{1}){\bf t}_{02}^{\prime }(s_{2})+\left\langle
\delta \varphi (s_{1})\delta \theta (s_{2})\right\rangle {\bf t}%
_{02}^{\prime }(s_{1}){\bf t}_{01}^{\prime }(s_{2}),  \nonumber
\end{align}
where ${\bf t}_{0i}{\bf t}_{0j}$ denotes the direct product of two vectors $%
{\bf t}_{0i}$ and ${\bf t}_{0j}$. The correlation functions of the Euler
angles that appear in the above expressions are given in Eq. (\ref{ang}). As
expected, the normalization condition for unit vectors, $\left\langle {\bf t}%
_{3}(s)\bullet {\bf t}_{3}(s)\right\rangle =1$, is satisfied up to terms of
second order in $\delta \eta $.

Using the equality 
\begin{equation}
\int_{0}^{s}ds_{1}\int_{0}^{s}ds_{2}f\left( 
{\displaystyle{s_{2}-s_{1} \over r}}%
\right) =2r^{2}\int_{0}^{s/r}du\left( 
{\displaystyle{s \over r}}%
-u\right) f\left( u\right) ,  \label{equal}
\end{equation}
valid for any even function $f(x)$, we obtain 
\begin{align}
\left\langle \left[ {\bf x}(s_{1})-{\bf x}\left( s_{2}\right) \right]
^{2}\right\rangle & =\int_{s_{1}}^{s_{2}}ds^{\prime
}\int_{s_{1}}^{s_{2}}ds^{\prime \prime }\left\langle {\bf t}_{3}(s^{\prime
})\bullet {\bf t}_{3}(s^{\prime \prime })\right\rangle  \nonumber \\
& =2r^{2}\left[ 1-\cos \left( 
{\displaystyle{s \over r}}%
\right) \right] -%
{\displaystyle{r^{3} \over \pi }}%
\left[ 
{\displaystyle{1 \over a_{\Vert }}}%
g_{\Vert }\left( 
{\displaystyle{s \over r}}%
\right) +%
{\displaystyle{1 \over a_{\bot }}}%
g_{\bot }\left( 
{\displaystyle{s \over r}}%
\right) +%
{\displaystyle{1 \over a_{3}}}%
g_{3}\left( 
{\displaystyle{s \over r}}%
\right) \right] ,  \label{x-x}
\end{align}
where $s=\left| s_{2}-s_{1}\right| ,\quad 0<s<2\pi r$. and where $a_{\bot
}^{-1}\ $is defined in Eq. (\ref{ainv}). The functions $g_{\Vert },$ $%
g_{\bot }$ and $g_{3}$ are given by 
\begin{align}
g_{\Vert }(x)& =2\int_{0}^{x}\left( x-u\right) \left[ f_{2}(0)-f_{2}\left(
u\right) \right] \cos udu  \nonumber \\
& =-\left( 1+\cos x\right) \left( \pi x-%
{\displaystyle{x^{2} \over 2}}%
\right) -\frac{1}{2}\left( 1+\cos x\right) ^{2}+2\left( \pi -x\right) \sin
x+2,  \nonumber \\
g_{\bot }(x)& =2\int_{0}^{x}\left( x-u\right) \left[ f_{1}(0)\cos
u-f_{1}\left( u\right) \right] du  \nonumber \\
& =-\frac{1}{2}\left( x\pi -%
{\displaystyle{x^{2} \over 2}}%
+1\right) \cos x+\frac{1}{2}\left( \pi -x\right) \sin x+\frac{1}{2},
\label{gi} \\
g_{3}(x)& =2\int_{0}^{x}\left( x-u\right) \left[ f_{3}(0)\cos u-f_{3}\left(
u\right) \right] du  \nonumber \\
& =-\frac{1}{2}\left( x\pi -%
{\displaystyle{x^{2} \over 2}}%
+3\right) \cos x+\frac{3}{2}\left( \pi -x\right) \sin x-x\pi +\frac{x^{2}}{2}%
+\frac{3}{2}.  \nonumber
\end{align}

For small $x\ll 1$ we have $g_{\Vert }(x)\simeq g_{\bot }(x)\simeq \pi
x^{3}/12$ and $g_{3}(x)\simeq x^{4}/32\ll g_{\Vert }(x)$. Combining these
expressions into Eq. (\ref{x-x}), we conclude that the lowest order
corrections to the straight line result, $\left\langle \left[ {\bf x}(s)-%
{\bf x}\left( 0\right) \right] ^{2}\right\rangle _{s/r\rightarrow 0}=s^{2}$,
depend only on the effective bending persistence length $2/\left(
a_{1}^{-1}+a_{2}^{-1}\right) $, in agreement with the wormlike chain model.
For general $s$ this correlator depends on all the bare persistence lengths, 
$a_{1},$ $a_{2}$ and $a_{3}$.

In Fig. 4 we plot the mean square distance between two points on the ring
contour, $\left\langle \left[ {\bf x(}s)-{\bf x(}0)\right] ^{2}\right\rangle 
$, as a function of $s$, in the interval $0\leq s\leq 2\pi r$. As expected,
it increases parabolically with $s$ (straight rod behavior for small $s$)
and exhibits a maximum at $s=\pi r$ (the maximum is determined by the
geometry of the undeformed ring). Fluctuations suppress this maximum in a
way that depends on the various rigidity parameters. Thus, decreasing the
twist rigidity, $a_{3}$, has a much smaller effect on the amplitude of the
maximum, than decreasing the bending rigidities $a_{1}$ or $a_{2}$. The
origin of this effect is that twist rigidity does not affect the spatial
conformations of a short segment of the ring that can be considered as a
nearly straight incompressible rod. Therefore, twist fluctuations affect
only the conformations of long segments, for which deviations from a
straight rod become significant (compare solid line and boxes in Fig. 4).

The radius of gyration is defined as 
\begin{equation}
R_{g}^{2}=\frac{1}{2\pi r}\oint ds\left[ {\bf x}(s)-\frac{1}{2\pi r}\oint
ds^{\prime }{\bf x}(s^{\prime })\right] ^{2}.  \label{Rg}
\end{equation}
Averaging this expression over fluctuations, we can express $\left\langle
R_{g}^{2}\right\rangle $ in terms of the two--point correlation function 
\begin{equation}
\left\langle R_{g}^{2}\right\rangle =\frac{1}{8\pi ^{2}r^{2}}\oint
ds_{1}\oint ds_{2}\left\langle \left[ {\bf x}(s_{1})-{\bf x}\left(
s_{2}\right) \right] ^{2}\right\rangle .  \label{Rgav}
\end{equation}
Using Eqs. (\ref{x-x}) and (\ref{gi}) we find 
\begin{equation}
\left\langle R_{g}^{2}\right\rangle =r^{2}\left[ 1-\left( \allowbreak
\allowbreak \frac{17}{8\pi }-\frac{\pi }{6}\right) 
{\displaystyle{r \over a_{\Vert }}}%
-\frac{3}{4\pi }%
{\displaystyle{r \over a_{\bot }}}%
-\left( \frac{7}{4\pi }-\frac{\pi }{6}\right) 
{\displaystyle{r \over a_{3}}}%
\right] .  \label{RG2}
\end{equation}
All the corrections to the unperturbed result ($r^{2}$) are negative, and we
conclude that fluctuations make the ring more compact. Since our weak
fluctuation approximation is only valid in the range $a_{i}\gg r,$ these
fluctuation corrections are rather small. Because of the small coefficient
in front of the $r/a_{3}$ term, the effect of twist fluctuations on the
radius of gyration is relatively weak, but fluctuations diverge and the
expression for the radius of gyration becomes unphysical in the limit of
vanishing twist rigidity, $a_{3}\rightarrow 0${\em . }

\section{Writhe fluctuations\label{WRITHE}}

The twist number $Tw$ associated with a configuration of the ring can be
expressed through the Euler angles, 
\begin{equation}
Tw=%
{\displaystyle{1 \over 2\pi }}%
\oint \omega _{3}(s)ds=%
{\displaystyle{1 \over 2\pi }}%
\int_{0}^{2\pi r}\left( 
{\displaystyle{d\psi  \over ds}}%
+\cos \theta 
{\displaystyle{d\varphi  \over ds}}%
\right) ds,  \label{Tw}
\end{equation}
where we used the definition of the rate of twist, $\omega _{3}(s)$, about
the tangent vector, in terms of the Euler angles, Eq. (\ref{o3}). In order
to understand the physical meaning of $\omega _{3}$, consider the variation
of the triad vector ${\bf t}_{1}$ (or{\bf \ }${\bf t}_{2}$) as one moves an
infinitesimal contour distance $ds$ along the centerline of the curved
filament. The projection of the vector ${\bf t}_{1}(s+ds)$ on the cross
section at $s$ (the plane normal to the tangent ${\bf t}_{3}(s)$), rotates
by an angle $\omega _{3}(s)ds$ compared to its original direction, ${\bf t}%
_{1}(s)$. Inspection of Eq. (\ref{Tw}) shows that this rotation consists of
two contributions. The first term corresponds to the contribution of a
straight filament of length $ds$ (the normal planes at points $s$ and $s+ds$
remain parallel to each other), whose cross section is twisted around the
centerline, by an angle $d\psi $. The second term, $\cos \theta d\varphi /ds$%
, arises due to the curvature of the centerline; since the cross sections at
points $s$ and $s+ds$ are, in general, tilted with respect to each other,
the projection of ${\bf t}_{1}(s+ds)$ on the cross section at $s$ will
rotate by $\cos \theta d\varphi $. Notice that because of the interplay of
the two effects, a curved filament can have zero twist even if $d\psi
/ds\neq 0$. This effect will be demonstrated in Section \ref{FDT} (see Fig.
6).

In addition to the twist of the filament that is closely associated with the
rotation of the cross section about a curved centerline, and can be defined
both locally (the twist ``density'' $\omega _{3}(s)$) and globally ($Tw$),
one can introduce an integral characteristic of the spatial configuration of
the centerline that reflects its tortuosity, known as writhe number. In
order to express the writhe number $Wr$ of a given configuration of the ring
in terms of Euler angles, one usually begins with the Fuller equation for
the writhe of a closed curve \cite{Fuller}: 
\begin{equation}
Wr=%
{\displaystyle{1 \over 2\pi }}%
\oint 
{\displaystyle{\left( {\bf t}_{03}{\bf \times t}_{3}\right) \bullet %
{\displaystyle{d \over ds}}\left( {\bf t}_{03}{\bf +t}_{3}\right)  \over 1+{\bf t}_{03}{\bf \cdot t}_{3}}}%
ds,  \label{wr1}
\end{equation}
where $\times $ and $\bullet $ denote respectively vector and scalar
products. In the above expression we made use of the fact that the writhe
number of a planar circular ring vanishes\cite{Tobias}. The above expression
is valid as long as $\left| {\bf t}_{03}\cdot {\bf t}_{3}(s)\right| <1$ in
the denominator, for all points $s$ on the contour of the ring. This
condition is satisfied in our work, since we only consider small
fluctuations about a planar undeformed ring that lies in the $xy$ plane.

A more physically transparent definition of writhe is based on the existence
of a topological invariant of a ring, called the linking number\cite
{Topology} $Lk$. The total rotation of the cross section as one moves around
the contour of the ring is given by $2\pi Lk$\ where the linking number 
\begin{equation}
Lk=Tw+Wr,  \label{Lk}
\end{equation}
does not depend on the conformation of the ring and is therefore a conserved
quantity. In the absence of spontaneous twist both the twist and the writhe
numbers of a planar circular ring vanish, and $Lk=Lk_{eq}=0$. In general,
since $Lk$\ is a constant for a given topology, $\delta Lk=\delta Wr+\delta
Tw=0$,\ and expanding the integrand in Eq. (\ref{Tw}) in small deviations of
the Euler angles from their spontaneous values in the unperturbed ring, the
deviations of writhe and twist can be expressed as 
\begin{equation}
\delta Wr=-\delta Tw=%
{\displaystyle{1 \over 2\pi }}%
\int_{0}^{2\pi r}%
{\displaystyle{d\delta \varphi  \over ds}}%
\delta \theta ds=\sum\nolimits_{n}in\tilde{\varphi}(n)\tilde{\theta}(-n),
\label{dWr}
\end{equation}
where we used $\oint d\delta \varphi =0$ and $\oint \delta \theta ds=0$ (see
Eq. (\ref{bound})). The last equality in Eq. (\ref{dWr}) was derived using
Eq. (\ref{exp}). Notice that the integrand in Eq. (\ref{dWr}) depends on the
product of $\delta \varphi $ and $\delta \theta ,$ and we conclude that
writhe deviations vanish both when fluctuations of the angles are confined
to the plane of the ring ($\delta \theta =0$), and when they are normal to
this plane ($\delta \varphi =0$).

Since $\left\langle \tilde{\varphi}(n)\tilde{\theta}(-n)\right\rangle $ is
an even function of $n$, multiplying by $n$ and summing over all positive
and negative integer values of $n$ yields 
\begin{equation}
\left\langle \delta Wr\right\rangle =\sum\nolimits_{n}in\left\langle \tilde{%
\varphi}(n)\tilde{\theta}(-n)\right\rangle =0.  \label{avWr}
\end{equation}

The dispersion of the writhe number is given by 
\begin{eqnarray}
\left\langle \delta Wr^{2}\right\rangle &=&\sum_{n\neq 0,\pm 1}Wr^{2}(n), 
\nonumber \\
Wr^{2}(n) &=&n^{2}\left[ \left\langle \tilde{\theta}(n)\tilde{\theta}%
(-n)\right\rangle \left\langle \tilde{\varphi}(n)\tilde{\varphi}%
(-n)\right\rangle -\left\langle \tilde{\varphi}(n)\tilde{\theta}%
(-n)\right\rangle ^{2}\right] ,  \label{dWr2}
\end{eqnarray}
where we excluded modes with $n=0$ and $n=\pm 1$, because of the boundary
conditions $\tilde{\theta}(0)=\tilde{\varphi}(1)=0$. Using Eqs. (\ref{ang})
for the correlation functions of Euler angles, we obtain the mean square
amplitude of writhe fluctuations at wavelength $r/n$, 
\begin{equation}
Wr^{2}(n)=%
{\displaystyle{r^{2} \over \pi ^{2}a_{1}a_{2}a_{3}}}%
{\displaystyle{A_{1}+a_{3}n^{2} \over \left( n^{2}-1\right) ^{2}}}%
.  \label{Wrn}
\end{equation}
Notice that the amplitude of writhe fluctuations diverges at $a_{3}=0$\ and
we conclude that twist rigidity plays an essential role in stabilizing the
contour of the ring against writhe fluctuations. The origin of this
divergence is the same as that of the correlator $\left\langle \delta \theta
(s)\delta \theta \left( 0\right) \right\rangle $\ in Eq. (\ref{ang}) and has
been discussed following Eq. (\ref{fi}).

For large wavevectors, $\left| n\right| \gg 1$, the mean square amplitude of
the n-th mode of writhe fluctuations depends only on the bending persistent
lengths, $a_{1}$\ and $a_{2}$. The physical reason is that on sufficiently
small scales, the filament behaves as a straight incompressible rod whose
properties do not depend on the twist persistence length $a_{3}$ (see
reference \cite{Mezard}). In the limit $A_{1}\ll a_{3}n^{2}$, the
writhe--writhe correlation function for a straight rod takes the form, 
\begin{equation}
Wr^{2}(q)=%
{\displaystyle{4 \over a_{1}a_{2}q^{2}}}%
,  \label{rod}
\end{equation}
where we defined the wavevector $q=2\pi n/r$. Eq. (\ref{rod}) is valid for
straight rods when $2\pi /q\ll a_{b}$, where the persistence length $a_{b}$
of the rod is defined by 
\begin{equation}
{\displaystyle{1 \over a_{b}}}%
=%
{\displaystyle{1 \over 2}}%
\left( 
{\displaystyle{1 \over a_{1}}}%
+%
{\displaystyle{1 \over a_{2}}}%
\right) .  \label{ab}
\end{equation}
The crossover to a long wavelength regime in which writhe modes become
affected by twist rigidity, takes place at a length scale $\xi _{t}=r\sqrt{%
a_{b}/a_{3}}$ and, therefore, such a regime exists for a ring of radius $r$
only if $a_{b}/a_{3}\leq 1$.\ As a consistency check, notice that the
straight rod case follows from the above expression for $\xi _{t}$ by
substituting $r=\infty $, and since $\xi _{t}$ diverges in this limit, Eq. (%
\ref{rod}) applies throughout the entire range of parameters.

Substituting Eq. (\ref{Wrn}) back in Eq. (\ref{dWr2}) yields 
\begin{equation}
\left\langle \delta Wr^{2}\right\rangle =\left( \frac{1}{6}+\frac{1}{8\pi
^{2}}\right) 
{\displaystyle{r^{2} \over a_{1}a_{2}}}%
+\left( \frac{1}{6}-\frac{11}{8\pi ^{2}}\right) 
{\displaystyle{r^{2} \over a_{\Vert }a_{3}}}%
.  \label{Wr3}
\end{equation}
Notice that $\left\langle \delta Wr^{2}\right\rangle \sim r^{2}$, in
agreement with the scaling estimates in reference \cite{Maggs}. Indeed,
since writhe is a quadratic form of $\delta \varphi $ and $\delta \theta $
(see Eq. (\ref{dWr})), each of which has typical fluctuations of $\sqrt{r/a}$
($a$ is a characteristic persistence length), the characteristic amplitude
of writhe fluctuations is $\delta Wr\approx r/a$.

The entire probability distribution of writhe can also be computed.
Beginning with the formal definition of this distribution 
\begin{equation}
P\left( \delta Wr\right) =\left\langle \delta \left[ \delta
Wr-\sum\limits_{n\neq 0,\pm 1}in\tilde{\varphi}(n)\tilde{\theta}(-n)\right]
\right\rangle ,  \label{P1}
\end{equation}
and using the exponential representation of the $\delta $-function, yields 
\begin{eqnarray}
P\left( \delta Wr\right) &=&\int_{-\infty }^{\infty }%
{\displaystyle{dk \over 2\pi }}%
e^{ik\delta Wr}\left\langle \exp \left[ k\sum\nolimits_{n}n\tilde{\varphi}(n)%
\tilde{\theta}(-n)\right] \right\rangle =  \nonumber \\
&&\int_{-\infty }^{\infty }%
{\displaystyle{dk \over 2\pi }}%
e^{ik\delta Wr}\prod_{n\neq 0,\pm 1}\sqrt{%
{\displaystyle{\det {\bf Q}(n) \over \det \left[ {\bf Q}(n)+kn{\bf Y}\right] }}%
},  \label{P2}
\end{eqnarray}
where the matrix ${\bf Y}$ is defined as 
\begin{equation}
{\bf Y}=%
{\displaystyle{r \over \pi }}%
\left( 
\begin{array}{ccc}
0 & -1 & 0 \\ 
1 & 0 & 0 \\ 
0 & 0 & 0
\end{array}
\right) .  \label{P}
\end{equation}
Calculating the corresponding determinants gives, 
\begin{equation}
P\left( \delta Wr\right) =\int_{-\infty }^{\infty }%
{\displaystyle{dk \over 2\pi }}%
{\displaystyle{e^{ik\delta Wr} \over \prod_{n=2}^{\infty }[1+Wr^{2}(n)k^{2}]}}%
.  \label{P3}
\end{equation}
This integral can be calculated by expanding the integrand into partial
fractions and we get 
\begin{eqnarray}
P\left( \delta Wr\right) &=&\sum_{n=2}^{\infty }\pi (n)%
{\displaystyle{1 \over 2Wr(n)}}%
\exp \left[ -%
{\displaystyle{|\delta Wr| \over Wr(n)}}%
\right] ,  \label{P4} \\
\pi (n) &=&\prod_{%
{k>1 \atop k\neq n}%
}\left[ 1-%
{\displaystyle{Wr^{2}(k) \over Wr^{2}(n)}}%
\right] ^{-1}.  \label{P4a}
\end{eqnarray}
Evaluating the product in Eq. (\ref{P4a}) yields, 
\begin{eqnarray}
\pi (n) &=&\left( -1\right) ^{n}%
{\displaystyle{\pi n^{2}\left( n^{2}+\alpha \right) \left( 1+\alpha \right) \sqrt{\alpha } \over 2\left( n^{2}-1\right) \sinh \left( \pi \sqrt{\alpha }\right) }}%
,  \label{calc} \\
\alpha (n) &\equiv &%
{\displaystyle{\left( n^{2}-2\right) A_{1}-a_{3} \over A_{1}+n^{2}a_{3}}}%
.  \nonumber
\end{eqnarray}

The above expression for $\pi (n)$ can be used to calculate all the even
moments of writhe fluctuations (odd moments vanish due to the radial
symmetry of the undeformed ring), 
\begin{equation}
\left\langle \delta Wr^{k}\right\rangle =k!\sum_{n=2}^{\infty }\pi
(n)Wr^{k}(n),\qquad \sum_{n=2}^{\infty }\pi (n)=1,\qquad k=2,4,\ldots
\label{mom}
\end{equation}
Since moments with $k>2$ do not vanish, it is obvious that the writhe
distribution is not Gaussian. Furthermore, inspection of Eq. (\ref{P4})
shows that the distribution has an exponential tail at large $\delta Wr$.
For strongly writhing rings, $\left\langle \delta Wr^{2}\right\rangle
^{1/2}\ll |\delta Wr|$, the $n=2$ term dominates the sum in Eq. (\ref{P4})
and the free energy $F=-k_{B}T\ln P\left( \delta Wr\right) $ is given by (up
to logarithmic corrections), 
\begin{equation}
{\displaystyle{F \over k_{B}T}}%
=%
{\displaystyle{\delta Wr \over Wr(2)}}%
\qquad \text{for}\qquad \left\langle \delta Wr^{2}\right\rangle ^{1/2}\ll
|\delta Wr|\ll 1.  \label{Fwr}
\end{equation}
The second inequality, $|\delta Wr|\ll 1$, follows from the assumption that
the deviations of Euler angles from their equilibrium values, are small.

The writhe distribution function can be written in the form 
\begin{equation}
P\left( \delta Wr\right) =%
{\displaystyle{\pi \sqrt{a_{1}a_{2}} \over r}}%
p\left( 
{\displaystyle{\pi \sqrt{a_{1}a_{2}} \over r}}%
\delta Wr,%
{\displaystyle{A_{1} \over a_{3}}}%
\right) .
\end{equation}
Plots of the dimensionless function $p(x,A_{1}/a_{3})$ for $A_{1}/a_{3}=0.1,$
$5$ and $20$ are shown in Fig. 5. As intuitively expected, the probability
of large writhe fluctuations (large $\pi \sqrt{a_{1}a_{2}}\delta Wr/r$)
decreases with increasing twist rigidity (decreasing $A_{1}/a_{3}$), but the
effect saturates for $A_{1}/a_{3}<0.1$. The shape of the curves bears close
resemblance to the results of recent computer simulations\cite{Beard}.

\section{Elastic response of the ring\label{FDT}}

According to the fluctuation--dissipation theorem, the correlation functions
of the Euler angles determine the elastic response of the ring to external
distributed torque ${\bf M}(s)$, applied along its contour. In the
following, we use this information in order to study the deformation of the
ring by external torques and forces. Since we are not interested in rigid
body rotation, we assume that the total torque on the ring vanishes, i.e., 
\begin{equation}
\oint ds{\bf M}(s)=0.  \label{total}
\end{equation}
The deviations of the Euler angles from their unperturbed values are given
by 
\begin{equation}
\delta \eta (s)=%
{\displaystyle{1 \over T}}%
\sum\nolimits_{\eta ^{\prime }}\oint ds^{\prime }\left\langle \delta \eta
(s)\delta \eta ^{\prime }(s^{\prime })\right\rangle M_{\eta ^{\prime
}}(s^{\prime }),  \label{fdt}
\end{equation}
where $\eta ,\eta ^{\prime }=\theta ,\varphi ,\psi $ and $M_{\eta ^{\prime
}} $ are the corresponding components of the torque. In order to calculate
the elastic response to external force, ${\bf F}(s)$, applied to the
centerline of the ring, we rewrite the work done by this force, $W=\oint ds%
{\bf F}\left( s\right) \bullet \delta {\bf x}(s)$, in the form 
\begin{equation}
W=\oint ds{\bf m}(s)\bullet \delta {\bf t}_{3}\left( s\right) ,\qquad \text{%
where}\qquad {\bf m}(s)=\int_{s}^{2\pi r}ds^{\prime }{\bf F}(s^{\prime }).
\label{ft3}
\end{equation}
Since we are not interested in translation of the ring as a whole, we assume
that the total force acting on the ring vanishes, $\oint ds{\bf F}(s)=0$,
which means that the function ${\bf m}(s)$ is continuous at $s=0$, i.e., $%
{\bf m}\left( 2\pi r\right) ={\bf m}(0)=0$. Using Eq. (\ref{dt3}) we can
recast the expression for the work done by the force, Eq. (\ref{ft3}), in
the form 
\begin{equation}
W=\oint ds\left\{ \left[ -m_{1}\left( s\right) \sin \left( s/r\right)
+m_{2}(s)\cos (s/r)\right] \delta \varphi (s)-m_{3}\left( s\right) \delta
\theta (s)\right\} .  \label{work}
\end{equation}
Inspection of this equation shows that in the presence of external force we
have to modify the expressions for the moments 
\begin{align}
M_{\varphi }(s)& \rightarrow M_{\varphi }(s)-m_{1}\left( s\right) \sin
(s/r)+m_{2}(s)\cos (s/r),  \nonumber \\
M_{\theta }(s)& \rightarrow M_{\theta }(s)-m_{3}(s),  \label{FDT1}
\end{align}
in Eq. (\ref{fdt}). The condition that the total torque due to the external
force vanishes, Eq. (\ref{total}), imposes additional conditions on the
force ${\bf F}(s)$. Upon some algebra, these conditions can be written in
the form, 
\begin{equation}
\int_{0}^{2\pi r}F_{t}(s)ds=\int_{0}^{2\pi r}sF_{3}(s)ds=0,  \label{FDT2}
\end{equation}
where $F_{t}(s)={\bf F}(s)\cdot {\bf t}_{03}(s)$ is the tangential component
of the force ${\bf F}$. Inspection of Eq. (\ref{FDT1}) shows that a small
force with $m_{1}(s)=-F_{r}(s)r\cos (s/r)$, $m_{2}(s)=-F_{r}(s)r\sin (s/r)$
and $m_{3}(s)=0$ does not deform the ring. From Eq. (\ref{ft3}) we find that 
$F_{r}(s)$ is the radial component of the force ${\bf F}(s)$, while its
tangential component is $F_{t}(s)=rd\left[ F_{r}(s)\right] /ds$. This
tensile force balances the contribution of variation of the radial force $%
F_{r}(s)$ along the contour of the ring and prevents buckling until a
critical value of the radial force is reached.

Equations (\ref{t30}) and (\ref{t123}) together with Eqs. (\ref{fdt}), (\ref
{FDT1}) and inextensibility condition, Eq. (\ref{inext}), determine the
conformation of the deformed ring, under the action of small external torque
and force. As an illustration, consider the deformation of the ring under
external forces $F$ applied to two opposite points $s=\pi /2$ and $s=3\pi /2$
on the ring contour, 
\begin{equation}
F_{1}(s)=F\left[ \delta \left( s-\pi /2\right) -\delta \left( s-3\pi
/2\right) \right] .  \label{F(s)}
\end{equation}
Using Eqs. (\ref{fdt}), (\ref{ft3}) and (\ref{FDT1}), we obtain the
following expressions for the resulting variations of Euler angles, 
\begin{eqnarray}
\delta \theta (s) &=&-F%
{\displaystyle{r^{2}\sin 2\psi _{0} \over 2\pi }}%
\left( 
{\displaystyle{1 \over a_{2}k_{B}T}}%
-%
{\displaystyle{1 \over a_{1}k_{B}T}}%
\right) h_{\theta }\left( \frac{s}{r}\right) ,  \nonumber \\
\delta \varphi (s) &=&F%
{\displaystyle{r^{2} \over \pi a_{\Vert }k_{B}T}}%
h_{\varphi }\left( \frac{s}{r}\right) ,  \label{var} \\
\delta \psi (s) &=&-F%
{\displaystyle{r^{2}\sin 2\psi _{0} \over 2\pi }}%
\left( 
{\displaystyle{1 \over a_{2}k_{B}T}}%
-%
{\displaystyle{1 \over a_{1}k_{B}T}}%
\right) h_{\psi }\left( \frac{s}{r}\right) ,  \nonumber
\end{eqnarray}
where 
\begin{eqnarray}
h_{\theta }(x) &=&%
{\displaystyle{dh_{\psi }(x) \over dx}}%
=\sum_{k=1}^{\infty }\frac{4k(-1)^{k}}{(4k^{2}-1)^{2}}\sin (2kx)=-%
{\displaystyle{\pi  \over 4}}%
x\cos x,  \nonumber \\
h_{\varphi }(x) &=&\sum_{k=1}^{\infty }\frac{(-1)^{k}}{k(4k^{2}-1)}\sin
(2kx)=x-%
{\displaystyle{\pi  \over 2}}%
\sin x,  \label{hi} \\
h_{\psi }(x) &=&-2\sum_{k=1}^{\infty }\frac{(-1)^{k}}{(4k^{2}-1)^{2}}\cos
(2kx)=\allowbreak 1-%
{\displaystyle{\pi  \over 4}}%
\cos x-%
{\displaystyle{\pi  \over 4}}%
x\sin x.  \nonumber
\end{eqnarray}
The above series are calculated in the interval $|x|<\pi /2$. Using the
periodicity condition, $h_{i}(x+\pi )=h_{i}(x)$, the functions $h_{i}(x)$
can be extended outside this interval. Inspection of Eq. (\ref{momin}) shows
that the persistence lengths $a_{i}$ are inversely proportional to
temperature. Since temperature enters Eq. (\ref{var}) only in combinations $%
a_{i}k_{B}T$, it cancels from the above expressions and can affect the
results only through temperature dependence of elastic moduli and moments of
inertia.

Eq. (\ref{var}) shows that the deformation of the ring under the action of
the force given in Eq. (\ref{F(s)}), does not depend on twist rigidity $%
a_{3} $. Therefore, such external forces do not produce any twist and can
only lead to bending of the ring. This result remains valid for more general
distributed forces on the centerline, provided they act only in the plane of
the undeformed ring. Inserting the expressions for the deviations of the
Euler angles, Eqs. (\ref{var}) and (\ref{hi}), into Eq. (\ref{domega}), we
find that $\delta \omega _{3}=0$ and consequently the variation of angle $%
\psi $\ can be expressed in terms of the conformation of the centerline
(angle $\theta $), as $rd\left[ \delta \psi (s)\right] /ds=\delta \theta (s)$%
. Since the sum of twist and writhe numbers is a topologically conserved
number, writhe is invariant under such deformations.

Figure 6 shows the effect of spontaneous (constant) angle of twist, $\psi
_{0}$, on the response of a ribbonlike ($a_{2}\gg a_{1}$) ring to
compressional forces applied at opposite points of the centerline. The
forces, shown by arrows, lie in the plane of the undeformed ring. In the
case of a ribbon with short axis lying in the plane of the undeformed ring, $%
\psi _{0}=\pi /2$, the ring remains planar in the course of deformation
(Fig. 6a). A ribbon with short axis lying normal to the plane of the ring, $%
\psi _{0}\rightarrow 0$, undergoes three dimensional deformation (Fig. 6b).
At first sight, Fig. 6b appears to suggest that the ring is twisted, in
contradiction with the previously made statement that its configuration is
twist--free. However, as is evident from the Eq. (\ref{domega}) for the
density of twist, $\delta \omega _{3}=d\delta \psi /ds-\delta \theta /r$,
and from the discussion in Section \ref{WRITHE}, the mathematical definition
of the twist of a filament with nonvanishing spontaneous curvature ($r\neq
\infty $) involves both the rotation of the cross section about the
centerline and the curvature of the centerline itself.{\em \ }The fact that
the two effects cancel exactly in Figs. 6a and 6b is a consequence of the
fact that the forces act entirely in the $xy$ plane and do not produce a
component of torque along the contour of the ring, that could give rise to
twist.

\section{Dynamics\label{DYNAMICS}}

Consider small instantaneous deviations $\delta {\bf x}(s,t)={\bf x}(s,t)-%
{\bf x}_{0}(s)$ of the centerline of the ring from its stress free position, 
${\bf x}_{0}(s)$. We express $\delta {\bf x}(s,t)$ in terms of its
projections on the triad vectors of the undeformed ring, ${\bf t}_{0k}(s)$, 
\begin{equation}
\delta {\bf x}(s,t){\bf =}\sum\nolimits_{k}\delta x_{k}(s,t){\bf t}_{0k}(s).
\label{dx}
\end{equation}
We proceed to write the Langevin equations that govern the dynamics of
fluctuations of the centerline, $\delta {\bf x}${\bf ,} and the dynamics of
angular fluctuations of the cross section about the centerline, $\delta \psi 
$. Some care should be exercised in deriving the Langevin force from the
expression for the elastic energy, Eq. (\ref{Fdef}), since up to this point
we have used the inextensibility constraint, Eq. (\ref{inext}). In order to
avoid complications associated with the introduction of rigid constraints 
\cite{Doi}, we replace the strict inextensibility condition, $d\delta
x_{3}/ds=\omega _{02}\delta x_{1}-\omega _{01}\delta x_{2}$ (see Appendix 
\ref{LANGEVEN}), by an energy penalty 
\begin{equation}
U_{ext}=%
{\displaystyle{k_{B}T \over 2r^{2}}}%
a_{ext}\int_{0}^{2\pi r}ds\left( \frac{d\delta x_{3}}{ds}-\omega _{02}\delta
x_{1}+\omega _{01}\delta x_{2}\right) ^{2}.  \label{ext}
\end{equation}
The persistence length $a_{ext}$ describes the rigidity of the filament with
respect to local compression and extension. The total elastic energy $%
U_{tot}=U+U_{ext}$ is the sum of contributions of bending and twist modes,
Eq. (\ref{Fr}), and extensional modes, Eq. (\ref{ext}). We will use the
above expression for the total energy of an extensible filament in the
Langevin equations, and take the limit of an inextensible filament ($%
a_{ext}/r\rightarrow \infty $) only in the end of the calculation.

The Langevin equations are 
\begin{eqnarray}
m\frac{d^{2}\delta {\bf x}(s,t)}{dt^{2}}+\varsigma \frac{d\delta {\bf x}(s,t)%
}{dt}+\frac{\delta U_{tot}}{\delta \lbrack \delta {\bf x}(s,t){\bf ]}} &=&%
{\bf f}(s,t),  \label{Lan1} \\
I\frac{d^{2}\delta \psi (s,t)}{dt^{2}}+\varsigma _{\psi }\frac{d\delta \psi
(s,t)}{dt}+\frac{\delta U_{tot}}{\delta \lbrack \delta \psi (s,t){\bf ]}}
&=&\xi _{\psi }(s,t).  \label{Lan2}
\end{eqnarray}
Here $m$ and $I$ are mass and moment of inertia (with respect to the
centerline) per unit length and $\varsigma $ and $\varsigma _{\psi }$ are
translational and angular friction coefficients. The
fluctuation--dissipation theorem relates the amplitudes of the random forces 
${\bf f}$ and $\xi _{\psi }$ to these friction coefficients, 
\begin{align}
\left\langle f_{i}(s,t)\right\rangle & =0,\qquad \left\langle
f_{i}(s,t)f_{j}(s^{\prime },t^{\prime })\right\rangle =2k_{B}T\varsigma
\delta _{ij}\delta (s-s^{\prime })\delta (t-t^{\prime }),  \label{fff} \\
\left\langle \xi _{\psi }(s,t)\right\rangle & =0,\qquad \left\langle \xi
_{\psi }(s,t)\xi _{\psi }(s^{\prime },t^{\prime })\right\rangle
=2k_{B}T\varsigma _{\psi }\delta (s-s^{\prime })\delta (t-t^{\prime }).
\label{ksiksi}
\end{align}
In writing the above equations we neglected hydrodynamic interactions and
therefore the treatment is analogous to the Rouse model of polymer solution
dynamics\cite{Doi}.

Using the relation between the deviations of the coordinates, $\delta {\bf x}
${\bf ,} and those of Euler angles, $\delta \theta $, $\delta \varphi $, and 
$\delta \psi $ (see Appendix \ref{LANGEVEN}), and neglecting rigid body
translation and rotation of the ring, we rewrite the above Langevin
equations in terms of the Fourier components of the deviations of Euler
angles from their equilibrium values (see Eq. (\ref{exp})), 
\begin{align}
\hat{\alpha}_{\eta }\tilde{\eta}(n,t)& =-L_{\eta }(n)%
{\displaystyle{\delta U \over \delta \tilde{\eta}(-n,t)}}%
+\tilde{\xi}_{\eta }(n,t),  \label{Lf1a} \\
\left\langle \tilde{\xi}_{\eta }(n,t)\right\rangle & =0,\qquad \left\langle 
\tilde{\xi}_{\eta }(n,t)\tilde{\xi}_{\eta ^{\prime }}(-n,t^{\prime
})\right\rangle =2k_{B}T\delta _{\eta \eta ^{\prime }}\varsigma _{\eta
}L_{\eta }(n)\delta (t-t^{\prime }).  \label{xi}
\end{align}
The time derivative operators $\hat{\alpha}_{\eta }$ associated with the
three Euler angles are, 
\begin{equation}
\hat{\alpha}_{\theta }=\hat{\alpha}_{\varphi }=\hat{\alpha}=m\frac{d^{2}}{%
dt^{2}}+\varsigma \frac{d}{dt}\quad \text{and \quad }\hat{\alpha}_{\psi 
\text{ }}=I\frac{d^{2}}{dt^{2}}+\varsigma _{\psi }\frac{d}{dt},
\label{alpha}
\end{equation}
where the corresponding friction coefficients are $\varsigma _{\theta
}=\varsigma _{\varphi }=\varsigma $ and $\varsigma _{\psi }$. The elastic
energy that appears in the Langevin equations (\ref{Lf1a}), is given in
terms of the amplitudes of the Fourier modes in Eq. (\ref{FreeEn}).
Conveniently, the matrix of kinetic coefficients ${\bf L}$ is diagonal in
the Euler angle representation, 
\begin{equation}
L_{\theta }(n)=n^{2}/r^{2},\quad L_{\varphi
}(n)=(n^{2}-1)^{2}/(n^{2}+1)r^{2}\quad \text{and}\quad L_{\psi }(n)=1/r^{2}.
\label{Ldiag}
\end{equation}

In the following we proceed to solve the Langevin equations and obtain
explicit expressions for the dynamic correlation function of writhe
fluctuations. We focus on this correlator since it is an integral
characteristic of the ring and is therefore simpler than the two--point
correlation functions of Euler angles, that depend on the separation between
the points. Although the general solution of the linear equations can be
obtained, we will assume (as it is often done in the literature\cite
{Maggs,GoGo-PRL-98,MoNe-Ma-98}) that the relaxation of the twist angle $\psi 
$ is much faster than that of the angles $\theta $ and $\varphi $.
Consequently, we can minimize the energy with respect to $\tilde{\psi}(n,t)$
and express it in terms of $\tilde{\theta}(n,t)$ and $\tilde{\varphi}(n,t)$.
With this substitution, the $(3\times 3)$ matrix ${\bf Q}(n)$ in $\tilde{%
\theta}(n,t)$, $\tilde{\varphi}(n,t),$ $\tilde{\psi}(n,t)$ space, Eq. (\ref
{F1}), is reduced to a $(2\times 2)$ matrix ${\bf Q}^{\prime }(n)$ in $%
\tilde{\theta}(n,t)$, $\tilde{\varphi}(n,t)$ space, 
\begin{equation}
{\bf Q}^{\prime }(n)=\frac{1}{a_{3}n^{2}+A_{1}}\left( 
\begin{array}{cc}
a_{3}A_{1}(n^{2}-1)^{2} & A_{3}a_{3}n^{2}(n^{2}-1) \\ 
A_{3}a_{3}n^{2}(n^{2}-1) & (a_{3}A_{2}n^{2}+a_{1}a_{2})n^{2}
\end{array}
\right)  \label{Q'}
\end{equation}

As shown in Appendix \ref{SOLUTION}, the solutions of the Langevin equations
can be expressed in terms of the eigenvalues $\Lambda _{1,2}(n)$ of the
matrix ${\bf P}(n)$ of the linear form 
\begin{equation}
L_{\eta }(n)%
{\displaystyle{\delta U \over \delta \tilde{\eta}(-n,t)}}%
=\sum\nolimits_{\eta ^{\prime }}P_{\eta \eta ^{\prime }}(n)\tilde{\eta}%
^{\prime }(n,t),\qquad \eta ,\eta ^{\prime }=\theta ,\varphi  \label{Peta}
\end{equation}
that appears in the Langevin equation, (\ref{Lf1a}). Explicit expressions
for these eigenvalues are given in Appendix \ref{SOLUTION}, Eq. (\ref{l12}).

Taking the Fourier transform of Eq. (\ref{ff1}) (Appendix \ref{SOLUTION})
with respect to the frequency $\omega $, we find the two--time correlation
function of the Fourier components of Euler angles, 
\begin{align}
\left\langle \tilde{\eta}(n,t)\tilde{\eta}^{\prime }(-n,0)\right\rangle & =%
\frac{1}{\Lambda _{2}(n)-\Lambda _{1}(n)}\left\{ \left[ \Lambda
_{2}(n)g_{1}(n,t)-\Lambda _{1}(n)g_{2}(n,t)\right] \times \right.  \nonumber
\\
& \left. \left\langle \tilde{\eta}(n)\tilde{\eta}^{\prime }(-n)\right\rangle
-k_{B}TL_{\eta }(n)\left[ g_{1}(n,t)-g_{2}(n,t)\right] \delta _{\eta \eta
^{\prime }}\right\} ,  \label{timecor}
\end{align}
where $g_{k}(n,t)$ describes temporal decay of correlations of normal modes
with wavevector $2\pi n/r$ ($g_{k}(n,0)=1$), and where $\left\langle \tilde{%
\eta}(n)\tilde{\eta}^{\prime }(-n)\right\rangle $ is the previously
calculated equal--time equilibrium correlation function (see section \ref
{ANGULAR}).

In the inertial limit (i.e., for modes with inertial time scale shorter than
viscous relaxation time), $4m\Lambda _{k}(n)>\varsigma ^{2}$, the function $%
g_{k}(n,t)$ describes damped oscillations with characteristic frequency $%
\omega _{k}(n)$, 
\begin{align}
g_{k}(n,t)& =\left[ \cos \omega _{k}(n)t+\frac{\varsigma }{2m\omega _{k}(n)}%
\sin \omega _{k}(n)t\right] \exp \left( -\frac{t\varsigma }{2m}\right) ,
\label{gk1} \\
\omega _{k}^{2}(n)& =\frac{\Lambda _{k}(n)}{m}-\frac{\varsigma ^{2}}{4m^{2}}.
\label{omega}
\end{align}
The characteristic relaxation time is $2m/\varsigma $, independent of the
wavelength of the mode. At short times, $t\ll m/\varsigma $, 
\begin{equation}
g_{k}(n,t)\simeq \cos \left[ \sqrt{\Lambda _{k}(n)/m}t\right] .
\label{small}
\end{equation}

In the dissipative limit, $4m\Lambda _{k}(n)<\varsigma ^{2}$, the function $%
g_{k}(n,t)$ describes pure decay of correlations, with characteristic times $%
\tau _{1}(n)$ and $\tau _{2}(n)$, 
\begin{align}
g_{k}(n,t)& =%
{\displaystyle{\tau _{1} \over \tau _{1}-\tau _{2}}}%
\exp \left( -\frac{t}{\tau _{1}}\right) +%
{\displaystyle{\tau _{2} \over \tau _{2}-\tau _{1}}}%
\exp \left( -\frac{t}{\tau _{2}}\right) ,  \label{gk2} \\
\tau _{1,2}(n)& =\frac{\varsigma }{2\Lambda _{k}(n)}\pm \sqrt{\frac{%
\varsigma ^{2}}{4\Lambda _{k}^{2}(n)}-\frac{m}{\Lambda _{k}(n)}}.
\label{tau}
\end{align}
In the limit of negligible inertia, $m\rightarrow 0$, we get $\tau
_{2}(n)\rightarrow 0$ and the relaxation can be described by simple
exponential decay, 
\begin{equation}
g_{k}(n,t)\simeq \exp \left[ -t\Lambda _{k}(n)/\varsigma \right] .
\label{expon}
\end{equation}

The dynamic writhe correlation function is derived in Appendix \ref{SOLUTION}%
: 
\begin{equation}
\left\langle \delta Wr(t)\delta Wr(0)\right\rangle =\sum\limits_{n\neq 0,\pm
1}Wr^{2}(n,t)=\sum\limits_{n\neq 0,\pm 1}Wr^{2}(n)g_{1}(n,t)g_{2}(n,t)
\label{WW}
\end{equation}
where the equilibrium mean squared amplitude of fluctuations of writhe, $%
Wr^{2}(n),$ is given in Eq. (\ref{Wrn}). In the limit of negligible inertia, 
$m\rightarrow 0$, we substitute Eq. (\ref{expon}) for $g_{k}(n,t)$ into Eq. (%
\ref{WW}), and find that the writhe correlation function for mode $n$ decays
exponentially with time 
\begin{equation}
Wr^{2}(n,t)=Wr^{2}(n)e^{-t/\tau (n)}.  \label{Wrnt}
\end{equation}
The characteristic decay time $\tau (n)$ is given by 
\begin{equation}
\tau (n)=\frac{\varsigma r^{3}}{\pi k_{B}T}\frac{n^{2}+1}{n^{2}(n^{2}-1)^{2}}%
\frac{a_{3}n^{2}+A_{1}}{(a_{1}+a_{2})a_{3}n^{2}+A_{1}a_{3}+a_{1}a_{2}}.
\label{taun}
\end{equation}
In the short wavelength limit, $n\gg 1$, only the sum of bending persistence
lengths, $a_{1}+a_{2}$, appears in $\tau (n)$. Indeed, on small scales the
filament behaves as a straight inextensible rod whose properties do not
depend on the twist persistence length $a_{3}$, or on the spontaneous twist
angle $\psi _{0}$.

In Figs. 7 and 8 we plot the writhe correlation function as a function of
time measured in units of $\varsigma r^{2}/\pi k_{B}T$, in the inertial
regime, for $2\pi mk_{B}T/\varsigma ^{2}r^{2}=10$. Its Fourier transform
plotted as a function of the frequency $\omega $ measured in units of $\pi
k_{B}T/\varsigma r^{2}$, is shown in the insert, on the upper right side of
the figure. In Fig. 7 the parameters correspond to a circular cross section
and identical persistence lengths, $a_{1}=a_{2}=a_{3}=2r$. Oscillatory decay
of writhe correlations as a function of time is observed, but the
correlation remains always positive. A small number of fundamental
frequencies can be detected in the oscillatory pattern, and identified with
peaks observed in the frequency spectrum. The amplitudes of these peaks
decrease monotonically with the frequency, and the largest peak is at $%
\omega =0$. The case of asymmetric cross section and dominant bending
rigidity, $a_{1}=a_{3}=2r$ and $a_{2}=20r$ ($\psi _{0}=\pi /4$) is shown in
Fig. 8. The correlation function decays rapidly to zero and, at later times,
oscillates between positive and negative values. Since for $t\ll m/\varsigma 
$, dissipation is negligible, the fast initial decay of correlations is the
result of dephasing of the oscillatory contributions of a large number of
modes. In the frequency domain, there is no peak at $\omega =0$ and the peak
amplitudes have nonmonotonic dependence on the frequency.

In Fig. 9 we plot the writhe correlation function in the dissipative regime
where inertia is negligible, $2\pi mk_{B}T/\varsigma ^{2}r^{2}=10^{-3}$, for
ribbonlike rings with different bending and twist rigidities. The amplitude
of writhe fluctuations is smallest for a ribbon whose shorter axis of
inertia is normal to the plane of the ring (see Fig. 6b). The amplitude
increases by more than a factor of two when the shorter axis of inertia lies
in the plane of the ring (see Fig 6a). Twist rigidity decreases the
fluctuation amplitude but the effect is rather weak. Since inertial
oscillations are completely suppressed in this overdamped regime, the
correlations decay monotonically with time. The curves exhibit fast short
time relaxation, followed by an exponential decay, in qualitative agreement
with numerical simulations reported in reference \cite{Beard}. An analytic
expression for the time dependence of the correlator at short times can be
derived (see Eq. (\ref{WrWr2}) in Appendix \ref{SOLUTION}). The predicted $%
\left\langle \left[ \delta Wr(t)-\delta Wr(0)\right] ^{2}\right\rangle
\propto $ $t^{1/4}$ dependence of the writhe correlation function and the
fact that it depends only on the bending rigidity ($a_{1}$ and $a_{2}$), are
direct consequences of the observation that at short times, the relaxation
is dominated by straight rod contributions to the spectrum ($\tau (q)\propto
1/q^{4}$). 

The above results can be directly applied to the study of deformation of
macroscopic rings by external forces and torques. Unlike the case of
microscopic filaments where dissipation dominates inertia and only
overdamped behavior is expected, inertial effects play an important role in
the dynamic response of macroscopic objects (for this reason they were
included in the preceding analysis). According to the
fluctuation--dissipation theorem, dynamic correlation functions of Euler
angles can be treated as generalized susceptibilities that determine the
response of the ring to externally applied torques and forces. The
time--dependent generalization of the static relation between deformation
(in terms of the deviation of the Euler angles from their equilibrium
values) and applied force, Eq. (\ref{fdt}), is 
\begin{equation}
\delta \eta (s,t)=%
{\displaystyle{1 \over T}}%
\int_{-\infty }^{t}dt^{\prime }\sum\nolimits_{\eta ^{\prime }}\oint
ds^{\prime }\frac{d}{dt^{\prime }}\left\langle \delta \eta (s,t)\delta \eta
^{\prime }(s^{\prime },t^{\prime })\right\rangle M_{\eta ^{\prime
}}(s^{\prime },t^{\prime }),  \label{fdtt}
\end{equation}
where the moments ${\bf M}$ due to external forces, are given by Eq. (\ref
{FDT1}). The relaxation of the deformation following the release of external
moments at time $t=0$, ${\bf M}(s,t)={\bf M}(s)\theta (-t)$, for $t>0$ is
given by 
\begin{equation}
\delta \eta (s,t)=%
{\displaystyle{1 \over T}}%
\sum\nolimits_{\eta ^{\prime }}\oint ds^{\prime }\left\langle \delta \eta
(s,t)\delta \eta ^{\prime }(s^{\prime },0)\right\rangle M_{\eta ^{\prime
}}(s^{\prime }).  \label{fdtt1}
\end{equation}

\section{Discussion\label{DISCUSS}}

In this paper we presented the statistical mechanics of fluctuating rings.
We derived analytical expressions for various static properties of such
rings, including two--point correlation functions of Euler angles,
correlation function of tangents to the ring, rms distance between points on
the ring contour, radius of gyration and probability distribution function
of writhe, as function of persistence lengths associated with bending and
twist deformations of the ring. We found that the amplitudes of fluctuations
of the Euler angles $\psi $\ and $\theta $\ diverge in the limit of
vanishing twist rigidity. We would like to emphasize that the situation
differs from the case of straight filaments for which the twist density $%
\delta \omega _{3}^{rod}=d\delta \psi /ds$\ depends only on the fluctuations
of the Euler angle $\psi .$\ For such filaments, vanishing twist rigidity ($%
a_{3}=0$) implies that there is no energy penalty for twisting the cross
section about the centerline, but the presence of bending rigidity ($%
a_{1},a_{2}\neq 0$) suffices to suppress spatial fluctuations of the
centerline about its straight stress free configuration. Thus, if we are
only interested in the statistical mechanics of the spatial conformations of
the centerline, accounting for bending rigidity suffices to provide an
accurate description of straight fluctuating filaments. For rings,
inspection of the elastic energy, Eq. (\ref{Fr}), shows that when $a_{3}=0$\
fluctuations with $d\delta \theta /ds+\delta \psi /r=0$\ have zero energy
cost and, since in the absence of twist rigidity the angle $\delta \psi $
can always adjust itself to satisfy the condition $\delta \psi =-r$ $d\delta
\theta /ds,$\ there is no elastic energy penalty for out--of--plane
fluctuations of the ring and the amplitude of such fluctuations diverges
even if the bending rigidity remains finite. Therefore, wormlike chain
theories in which only bending rigidity is taken into account, can not model
the spatial conformation of fluctuating rings.

We found that a crossover length scale $\xi _{t}=r\sqrt{a_{3}/a_{b}}$
exists, below which straight rod behavior dominates and writhe of the
centerline and twist of the cross section about it are decoupled, and above
which spontaneous curvature becomes important and twist affects the three
dimensional configurations of the centerline of the ring. In this context we
would like to propose the as yet unproven but plausible conjecture, that the
existence of this crossover does not depend on the topology of the ring and
is characteristic of filaments with spontaneous curvature in their stress
free state.

Although the main focus of this work is on the statistical mechanics of
fluctuating rings, we used the fluctuation--dissipation theorem in order to
predict mechanical response to external torques and forces, and showed that
the deformation of ribbonlike rings depends in an essential way on the
orientation of the cross section in the stress free state. Finally, we
derived the Langevin equations that govern the dynamics of fluctuating
rings, and calculated the two--time correlation function of writhe
fluctuations. Depending on the values of the parameters, one can move from
an inertial regime where relaxation is accompanied by temporal oscillations,
to a non--oscillatory, purely dissipative regime. In the dissipation
dominated range, the relaxation at short times is determined by bending
rigidity only. This agrees with the expectation that short time relaxation
is dominated by small scale, straight rod behavior. At longer times, the
decay of writhe correlations depends on both bending and twist rigidities.
While inertial effects are not expected to be important for microscopic
objects such as small plasmids, our dynamic response functions can describe
the relaxation of rings of arbitrary mass and size, following the cessation
of externally applied forces and torques.

We would like to comment on the limitations of the approach presented in
this paper. The domain of applicability of our theory is limited to the weak
fluctuation regime, in the sense that the deviations of the Euler angles
from their values in the undeformed ring, must be sufficiently small.
Although, in principle, our general formalism is applicable to rings with
arbitrary spontaneous twist in their stress free reference state, the
analysis of this problem meets with considerable mathematical difficulties
and is the subject of ongoing work. Finally, we would like to emphasize that
since our theory is based on the linear theory of elasticity of thin rods,
all persistence lengths and radii of curvature are assumed to be much larger
than the diameter of the filament that serves as a small scale cutoff.
Consideration of microscopic physics on length scales smaller than this
diameter requires the introduction of additional model assumptions (see
references \cite{Kremer,Haijun,Liverpool}) and is beyond the scope of this
paper.

After this work was submitted for publication we learned about a related
study of thermal fluctuations in DNA plasmids in which the writhe
distribution function was also calculated\cite{miniplasmids}. This work is
complementary to ours: while we assume that the equilibrium stress free
state of our filament is that of a planar untwisted circular ring, reference 
\cite{miniplasmids} deals with filaments with straight untwisted stress free
state. Conceptually, the ring is then formed by bringing the ends together
and sealing them, with or without the addition of twist. Since such rings
have locked--in internal stresses, the two procedures are nonequivalent in
general.

{\Large {\bf Acknowledgment}}

We would like to thank A. Grosberg, A. Maggs, T. Schlick and I. Tobias for
helpful correspondence, and D. Kessler for valuable comments on the
manuscript. YR acknowledges support by a grant from the Israel Science
Foundation.

\bigskip

\appendix

\section{Derivation of Langevin equations\label{LANGEVEN}}

For small deviations from the stress free state, the variation of triad of
vectors can be written as 
\begin{eqnarray}
\delta {\bf t}_{1} &=&-\left( \delta \theta \cos \psi _{0}+\delta \varphi
\sin \psi _{0}\right) {\bf t}_{03}+\delta \psi {\bf t}_{02},  \nonumber \\
\delta {\bf t}_{2} &=&\left( \delta \theta \sin \psi _{0}-\delta \varphi
\cos \psi _{0}\right) {\bf t}_{03}-\delta \psi {\bf t}_{02},  \label{dti} \\
\delta {\bf t}_{3} &=&-\left( \delta \theta \sin \psi _{0}-\delta \varphi
\cos \psi _{0}\right) {\bf t}_{02}+\left( \delta \theta \cos \psi
_{0}+\delta \varphi \sin \psi _{0}\right) {\bf t}_{01},  \nonumber
\end{eqnarray}
where the vectors ${\bf t}_{0i}$ are defined in Eq. (\ref{t0}). Substituting
Eqs. (\ref{dti}) into the inextensibility condition, Eq. (\ref{inext}), we
obtain the following equations for the deviation $\delta {\bf x}(s)$ of the
position vector of a point $s$ on the ring contour, from its value in the
undeformed state, 
\begin{eqnarray}
\frac{d\delta x_{3}}{ds}-\omega _{02}\delta x_{1}+\omega _{01}\delta x_{2}
&=&0,  \label{x3} \\
-\frac{d\delta x_{2}}{ds}+\omega _{01}\delta x_{3}-\omega _{03}\delta x_{1}
&=&\delta \theta \sin \psi _{0}-\delta \varphi \cos \psi _{0},  \label{x2} \\
\frac{d\delta x_{1}}{ds}+\omega _{02}\delta x_{3}-\omega _{03}\delta x_{2}
&=&\delta \theta \cos \psi _{0}+\delta \varphi \sin \psi _{0}.  \label{x1}
\end{eqnarray}
Eq. (\ref{x3}) is the linearized form of the inextensibility condition, and
Eqs. (\ref{x2}) and (\ref{x1}) relate the deviations of Euler angles to
those of spatial positions.

Fourier transforming Eqs. (\ref{Lan1}) and (\ref{Lan2}), yields the Langevin
equations for the Fourier components ${\bf \tilde{x}}\left( n\right) $ and $%
\tilde{\psi}(n)$, 
\begin{align}
\hat{\alpha}{\bf \tilde{x}}(n,t)+\frac{\delta U}{\delta {\bf \tilde{x}}(-n,t)%
}+\mu (n,t){\bf c}^{\ast }(n)& ={\bf \tilde{f}}(n,t),  \label{L11} \\
\hat{\alpha}_{\psi }\tilde{\psi}(n,t)+\frac{\delta U}{\delta \tilde{\psi}%
(-n,t)}& =\tilde{\xi}_{\psi }(n,t),  \label{L12}
\end{align}
where $\hat{\alpha}$ and $\hat{\alpha}_{\psi }$ are defined in Eq. (\ref
{alpha}) and 
\begin{equation}
\mu (n,t)=\frac{a_{ext}k_{B}T}{r^{2}}{\bf \tilde{x}}(n,t)\cdot {\bf c}%
(n),\quad \quad {\bf c}(n)={\bf (}-\sin \psi _{0},-\cos \psi _{0},in).
\label{mu}
\end{equation}
In the limit $a_{ext}\longrightarrow \infty $, $\mu (n,t)$ can be considered
as a Lagrange multiplier that accounts for the inextensibility condition,
Eq. (\ref{x3}) or, equivalently, for its Fourier transform, ${\bf \tilde{x}}%
(n,t)\cdot {\bf c}(n)=0$. The correlators of random forces in Eqs. (\ref{L11}%
) and (\ref{L12}), take the form, 
\begin{align}
\left\langle \tilde{f}_{i}(n,t)\right\rangle & =0,\qquad \left\langle \tilde{%
f}_{i}(n,t)\tilde{f}_{j}(-n,t^{\prime })\right\rangle =2\varsigma T\delta
_{ij}\delta (t-t^{\prime }),  \label{fifi} \\
\left\langle \tilde{\xi}_{\psi }(n,t)\right\rangle & =0,\qquad \left\langle 
\tilde{\xi}_{\psi }(n,t)\tilde{\xi}_{\psi }(-n,t^{\prime })\right\rangle
=2\varsigma _{\psi }T\delta (t-t^{\prime }).  \label{xixi}
\end{align}
Using Fourier transforms of Eqs. (\ref{x3}) -- (\ref{x1}) we rewrite the
Langevin equations (\ref{L11}) -- (\ref{xixi}) in terms of Euler angles,
Eqs. (\ref{Lf1a}) and (\ref{xi}).

\section{Solution of Langevin equations\label{SOLUTION}}

In order to find the solution of the Langevin equations, we first calculate
the eigenvalues and eigenfunctions of the matrix (see Eq. (\ref{Peta}) for
its definition), 
\begin{equation}
{\bf P}(n)=%
{\displaystyle{\pi k_{B}T \over r^{3}}}%
{\displaystyle{n^{2}(n^{2}-1) \over a_{3}n^{2}+A_{1}}}%
\left( 
\begin{array}{cc}
A_{1}a_{3}(n^{2}-1) & A_{3}a_{3}n^{2} \\ 
A_{3}a_{3}\frac{(n^{2}-1)^{2}}{n^{2}+1} & \left(
A_{2}a_{3}n^{2}+a_{1}a_{2}\right) \frac{n^{2}-1}{n^{2}+1}
\end{array}
\right)  \label{Pn}
\end{equation}
$\allowbreak $Eigenvalues of this matrix have the form 
\begin{align}
\Lambda _{1,2}(n)& =\frac{\pi k_{B}T}{2r^{3}}\frac{n^{2}(n^{2}-1)^{2}}{%
n^{2}+1}\frac{(a_{1}+a_{2})a_{3}n^{2}+A_{1}a_{3}+a_{1}a_{2}\pm \Delta }{%
a_{3}n^{2}+A_{1}},  \label{l12} \\
\Delta ^{2}&
=[(a_{1}-a_{2})a_{3}n^{2}+A_{1}a_{3}-a_{1}a_{2}]^{2}+4n^{2}a_{2}a_{3}(a_{1}-a_{2})(\allowbreak a_{1}-a_{3})\sin ^{2}\psi _{0}.
\nonumber
\end{align}
The eigenvalues $\Lambda _{k}(n)$ vanish when $n=0,1$. These modes are
associated with rigid body rotations of the ring and are not considered
further below. In the limit $|n|\gg 1$ both eigenvalues increase with the
fourth power of $n$, $\Lambda _{k}(n)\simeq \pi k_{B}Ta_{k}n^{4}/r^{3}$.
This $q^{4}$ dependence of the eigenvalues ($q=2\pi n/r$ is the wavevector
corresponding to the $n-$th mode) is characteristic of bending fluctuations
of straight rods, in accord with the expectation that small--scale
fluctuations of a ring are indistinguishable from those of a straight rod.

The matrix ${\bf P}(n)$ becomes diagonal (and the angles $\theta $ and $%
\varphi $ become decoupled), in the case of a circularly symmetric cross
section ($A_{3}=0$), when $\psi _{0}=0$ or $\pi /2$, and in the limit $%
a_{3}\rightarrow 0$. In all of these cases the mode $\Lambda _{1}(n)$
describes both fluctuations perpendicular to the plane of the ring and twist
fluctuations, and the mode $\Lambda _{2}(n)$ describes fluctuations in the
plane of the ring. Since $\Lambda _{1}(n)\rightarrow 0$ when $%
a_{3}\rightarrow 0$, twist fluctuations destroy the circular shape of the
ring in the wormlike chain model, where twist rigidity is not taken into
account.

The eigenvalues take a particularly simple form for $\psi _{0}=0$, 
\begin{equation}
\Lambda _{1}(n)=\frac{\pi k_{B}T}{r^{3}}\frac{n^{2}(n^{2}-1)^{2}a_{1}a_{3}}{%
a_{3}n^{2}+a_{1}},\qquad \Lambda _{2}(n)=\frac{\pi k_{B}T}{r^{3}}\frac{%
n^{2}(n^{2}-1)^{2}a_{2}}{n^{2}+1}.  \label{l12n}
\end{equation}

Since the matrix ${\bf P}(n)$, Eq. (\ref{Pn}), is not symmetric, it has
different right and left eigenfunctions. We denote the right eigenfunctions,
corresponding to eigenvalues $\Lambda _{k}(n)$, by $\bar{\eta}_{k}(n)=\{\bar{%
\theta}_{k}(n),\bar{\varphi}_{k}(n)\}$, (where $k=1,2$). Left eigenfunctions
can be written as $\bar{\eta}_{k}(-n)L_{\eta }^{-1}(n)$. Since the matrix $%
{\bf P}(n)$ is real, we have $\bar{\eta}_{k}(-n)=\bar{\eta}_{k}^{\ast }(n)$.
For each $n\neq 0,\pm 1$, the above eigenfunctions are normalized by
conditions: 
\begin{equation}
\sum\nolimits_{\eta }L_{\eta }^{-1}(n)\bar{\eta}_{k}(n)\bar{\eta}_{l}\left(
-n\right) =\delta _{kl},\qquad \sum\nolimits_{k}\bar{\eta}_{k}(n)\bar{\eta}%
_{k}^{^{\prime }}\left( -n\right) =L_{\eta }(n)\delta _{\eta \eta ^{\prime
}}.  \label{norm1}
\end{equation}
Expanding the matrix ${\bf P}(n)$ over its eigenfunctions we find: 
\begin{equation}
P_{\eta \eta ^{\prime }}(n)=\sum\nolimits_{k}\Lambda _{k}(n)\bar{\eta}_{k}(n)%
\bar{\eta}_{k}^{\prime }(-n)L_{\eta ^{\prime }}^{-1}(n).  \label{Pff}
\end{equation}

The solution of Eq. (\ref{Lf1a}) can be found by Fourier transforming it
with respect to the time $t$, and substituting Eqs. (\ref{Peta}) and (\ref
{Pff}). This yields 
\begin{align}
\tilde{\eta}_{\omega }(n)& =\sum_{k}\frac{\bar{\eta}_{k}(n)}{\tilde{\alpha}%
_{\omega }+\Lambda _{k}(n)}\sum_{\eta ^{\prime }}\bar{\eta}_{k}^{\prime }(n)%
\tilde{\xi}_{\eta ^{\prime }\omega }(n),  \label{sol1} \\
\tilde{\alpha}_{\omega }& =-m\omega ^{2}+i\varsigma \omega ,  \label{alpha1}
\end{align}
where the correlators of the Gaussian random force are 
\begin{equation}
\left\langle \tilde{\xi}_{\eta \omega }(n)\right\rangle =0,\qquad
\left\langle \tilde{\xi}_{\eta \omega }(n)\tilde{\xi}_{\eta ^{\prime
}-\omega }(-n)\right\rangle =2k_{B}T\varsigma L_{\eta }\delta _{\eta \eta
^{\prime }}.  \label{noise}
\end{equation}

Calculating the correlation functions of Euler angles, we get 
\begin{equation}
\left\langle \tilde{\eta}_{\omega }(n)\tilde{\eta}_{-\omega }^{\prime
}(-n)\right\rangle =2k_{B}T\varsigma \sum\nolimits_{k}\frac{\bar{\eta}_{k}(n)%
\bar{\eta}_{k}^{\prime }(-n)}{\left| \tilde{\alpha}_{\omega }+\Lambda
_{k}(n)\right| ^{2}},  \label{ff1}
\end{equation}
where $\eta ,\eta ^{\prime }=\theta ,\varphi $. In the time domain this
gives the following expression for the dynamic correlation functions of the
Fourier transforms of Euler angles, 
\begin{equation}
\left\langle \tilde{\eta}(n,t)\tilde{\eta}^{\prime }(-n,0)\right\rangle
=k_{B}T\sum\nolimits_{k}\frac{\bar{\eta}_{k}(n)\bar{\eta}_{k}^{\prime }(-n)}{%
\Lambda _{k}(n)}g_{k}(n,t),  \label{ff2}
\end{equation}
where 
\begin{equation}
g_{k}(n,t)=%
{\displaystyle{2\varsigma \Lambda _{k}(n) \over \pi }}%
\int_{0}^{\infty }%
{\displaystyle{d\omega \cos \omega t \over [\Lambda _{k}(n)-m\omega ^{2}]^{2}+\varsigma ^{2}\omega ^{2}}}%
.  \label{gknt}
\end{equation}
The function $g_{k}(n,t)$ describes temporal decay of correlations of normal
modes, with wavevector $2\pi n/r$. One can verify that the for $t=0$,
integration gives $g_{k}(n,0)=1$, independent of $m$ and $\varsigma $.
Instead of calculating the eigenfunctions $\bar{\eta}_{k}(n)$, we notice
that the combinations $\bar{\eta}_{k}(n)\bar{\eta}_{k}^{\prime }(-n)/\Lambda
_{k}(n)$ in the above expression, can be evaluated using the previously
derived expressions for the equilibrium (equal time) correlators $%
\left\langle \tilde{\eta}(n)\tilde{\eta}^{\prime }(-n)\right\rangle $, and
the normalization conditions, Eq. (\ref{norm1}). Substituting these
expressions into Eq. (\ref{ff2}), we arrive at Eq. (\ref{timecor}).

We now turn to writhe fluctuations (see Eq. (\ref{dWr})), 
\begin{equation}
\delta Wr(t)=-\sum\nolimits_{n}in\int \frac{d\omega }{2\pi }\tilde{\varphi}%
_{\omega }(n)e^{i\omega t}\int \frac{d\omega ^{\prime }}{2\pi }\tilde{\theta}%
_{\omega ^{\prime }}(-n)e^{i\omega ^{\prime }t},  \label{dWrr}
\end{equation}
and proceed to calculate the dynamic correlation function of these
fluctuations, 
\begin{equation}
\left\langle \delta Wr(t)\delta Wr(0)\right\rangle =\sum\limits_{n\neq 0,\pm
1}Wr^{2}(n,t),  \label{summode}
\end{equation}
where the contribution of mode $n$ is 
\begin{equation}
\begin{array}{c}
Wr^{2}(n,t)=k_{B}^{2}T^{2}n^{2}%
\displaystyle\int %
{\displaystyle{d\omega  \over 2\pi }}%
\displaystyle\int %
{\displaystyle{d\omega ^{\prime } \over 2\pi }}%
\cos [(\omega +\omega ^{\prime })t]\times \\ 
\left[ \left\langle \tilde{\theta}_{\omega }(-n)\tilde{\theta}_{-\omega
}(n)\right\rangle \left\langle \tilde{\varphi}_{\omega ^{\prime }}(n)\tilde{%
\varphi}_{-\omega ^{\prime }}(-n)\right\rangle -\left\langle \tilde{\theta}%
_{\omega }(-n)\tilde{\varphi}_{-\omega }(n)\right\rangle \left\langle \tilde{%
\theta}_{-\omega ^{\prime }}(n)\tilde{\varphi}_{\omega ^{\prime
}}(-n)\right\rangle \right] .
\end{array}
\label{Wr1}
\end{equation}
Substituting Eq. (\ref{ang}) for the correlation functions of Euler angles
yields, 
\begin{align}
Wr^{2}(n,t)& =k_{B}^{2}T^{2}n^{2}\sum\limits_{kk^{\prime }}\frac{%
g_{k}(n,t)g_{k^{\prime }}(n,t)}{\Lambda _{k}(n)\Lambda _{k^{\prime }}(n)}%
\times  \nonumber \\
& \left[ \bar{\theta}_{k}(n)\bar{\theta}_{k}(-n)\bar{\varphi}_{k^{\prime
}}(n)\bar{\varphi}_{k^{\prime }}(-n)-\bar{\theta}_{k}(n)\bar{\varphi}_{k}(-n)%
\bar{\varphi}_{k^{\prime }}(n)\bar{\theta}_{k^{\prime }}(-n)\right] .
\label{wr2nt}
\end{align}
The only nonvanishing contributions to the sum in Eq. (\ref{wr2nt}) are $%
k=1,k^{\prime }=2$ and $k=2,k^{\prime }=1$, and both have the same value. As
a result, we find that the contribution of the $n-$th mode to the dynamic
correlation function can be recast in the form 
\begin{equation}
Wr^{2}(n,t)=Wr^{2}(n)g_{1}(n,t)g_{2}(n,t),  \label{Wr2}
\end{equation}
where $Wr^{2}(n)$ is the mean squared amplitude of the $n-$th mode of writhe
fluctuations in equilibrium, calculated earlier in Eq. (\ref{Wrn}).

Finally, we would like to comment on the short time behavior of the
correlation function in the dissipative regime. Using Eq. (\ref{Wrnt}) we
find 
\begin{equation}
\left\langle \left[ \delta Wr(t)-\delta Wr(0)\right] ^{2}\right\rangle =%
\frac{4r^{2}}{\pi ^{2}a_{1}a_{2}a_{3}}\sum_{n=2}^{\infty }\frac{%
A_{1}+a_{3}n^{2}}{(n^{2}-1)^{2}}\left[ 1-e^{-t/\tau (n)}\right] .
\label{WrWr}
\end{equation}
At small times, $t\ll \tau (2)$, this sum is dominated by terms with $n\gg 1$%
. Replacing the sum by an integral we find 
\begin{equation}
\left\langle \left[ \delta Wr(t)-\delta Wr(0)\right] ^{2}\right\rangle =%
\frac{4r^{2}}{\pi ^{2}a_{1}a_{2}}\allowbreak \Gamma \left( \frac{3}{4}%
\right) \left[ \frac{\pi k_{B}T}{\varsigma r^{3}}(a_{1}+a_{2})t\right]
^{1/4},  \label{WrWr2}
\end{equation}
where $\Gamma $ is the gamma function. The characteristic relaxation rate $%
\pi k_{B}T(a_{1}+a_{2})/\varsigma r^{3}$ depends only on the bending
rigidity of the ring.

\newpage

{\LARGE {\bf Figure captions}}

{\bf Fig. 1:} Plots of two--point correlation functions of Euler angles $%
\left\langle \delta \eta (s)\delta \eta ^{\prime }(0)\right\rangle $ vs. the
contour distance between the points, $s$, in the interval $0\leq s\leq 2\pi
r $: $\left\langle \delta \theta (s)\delta \theta (0)\right\rangle $
(cross), $\left\langle \delta \varphi (s)\delta \varphi (0)\right\rangle $
(diamond), $\left\langle \delta \psi (s)\delta \psi (0)\right\rangle $
(circle) and $\left\langle \delta \theta (s)\delta \psi (0)\right\rangle $
(solid line). The parameters are $\psi _{0}=0$ and $a_{1}=a_{2}=10r,$ $%
a_{3}=100r$.

{\bf Fig. 2:} Plots of two--point correlation functions of Euler angles $%
\left\langle \delta \eta (s)\delta \eta ^{\prime }(0)\right\rangle $ vs. the
contour distance between the points, $s$, in the interval $0\leq s\leq 2\pi
r $: $\left\langle \delta \theta (s)\delta \theta (0)\right\rangle $
(cross), $\left\langle \delta \psi (s)\delta \psi (0)\right\rangle $(circle)
and $\left\langle \delta \theta (s)\delta \psi (0)\right\rangle $ (solid
line). The parameters are $\psi _{0}=0$ and $a_{1}=a_{2}=10r,$ $a_{3}=r$.

{\bf Fig. 3:} Plots of nondiagonal two--point correlation functions of Euler
angles $\left\langle \delta \eta (s)\delta \eta ^{\prime }(0)\right\rangle $
vs. the contour distance between the points, $s$, in the interval $0\leq
s\leq 2\pi r$: $\left\langle \delta \theta (s)\delta \varphi
(0)\right\rangle $ (cross), $\left\langle \delta \varphi (s)\delta \psi
(0)\right\rangle $ (circle) and $\left\langle \delta \theta (s)\delta \psi
(0)\right\rangle $ (solid line). The parameters are $\psi _{0}=\pi /4$ and $%
a_{1}=10r,a_{2}=100r,$ $a_{3}=10r$.

{\bf Fig. 4:} Plot of dimensionless rms distance between points on the ring
contour $\left\langle \left[ {\bf x(}s)-{\bf x(}0)\right] ^{2}\right\rangle
/r^{2}$ vs. the contour distance between the points, $s$, in the interval $%
0\leq s\leq 2\pi r$. The parameters are $\psi _{0}=0$ and: $%
a_{1}=a_{2}=a_{3}=10r$ (solid line), $a_{1}=a_{2}=10r,$ $a_{3}=r$ (box), $%
a_{1}=a_{2}=r,$ $a_{3}=10r$ (cross) and $a_{1}=10r,a_{2}=a_{3}=r$ (diamond).

{\bf Fig. 5:} Plot of probability distribution function of writhe $%
p(x,A_{1}/a_{3})$ vs. $x=\left( \pi \sqrt{a_{1}a_{2}}/r\right) \delta Wr$
for $A_{1}/a_{3}$ $=$ $0.1$ (solid line), $5$ (cross) and $20$ (box).

{\bf Fig. 6:} Plots of deformation of ribbonlike rings (with $%
a_{2}/a_{1}=10^{4}$) by compressional forces (see arrows). a) $\psi _{0}=\pi
/2$ and b) $\psi _{0}=10^{-3}$.

{\bf Fig. 7:} Plot of dynamic correlation function of writhe fluctuations $%
\left\langle \delta Wr(t)\delta Wr(0)\right\rangle $ vs. time $t$ (in units
of $\varsigma r^{2}/\pi k_{B}T$), for a ring with a circular cross section
and persistence lengths $a_{1}=a_{2}=a_{3}=2r,$ in the inertial range $2\pi
mk_{B}T/\varsigma ^{2}r^{2}=10$. Plot of the Fourier transform of the
correlation functions vs. frequency $\omega $ (in units of $\pi
k_{B}T/\varsigma r^{2}$) is shown as an insert in the upper right hand side
of the figure.

{\bf Fig. 8.} Plot of dynamic correlation function of writhe fluctuations $%
\left\langle \delta Wr(t)\delta Wr(0)\right\rangle $ vs. time $t$ (in units
of $\varsigma r^{2}/\pi k_{B}T$), for a ribbonlike ring with persistence
lengths $a_{1}=a_{3}=2r$ and $a_{2}=20r$ ($\psi _{0}=\pi /4$), in the
inertial range $2\pi mk_{B}T/\varsigma ^{2}r^{2}=10$. Plot of the Fourier
transform of the correlation functions vs. frequency $\omega $ (in units of $%
\pi k_{B}T/\varsigma r^{2}$) is shown as an insert in the upper right hand
side of the figure.

{\bf Fig. 9}. Plot of dynamic correlations function of writhe fluctuations $%
\left\langle \delta Wr(t)\delta Wr(0)\right\rangle $ vs. time $t\ $(in units
of $\varsigma r^{2}/\pi k_{B}T$), for ribbonlike rings in the dissipative
range $2\pi mk_{B}T/\varsigma ^{2}r^{2}=10^{-3}$: The parameters are $\psi
_{0}=0$ and $a_{1}=2r$, $a_{2}=20r,\,a_{3}=2r$ (cross), $a_{1}=20r$, $%
a_{2}=2r,\,a_{3}=2r$ (circle) and $a_{1}=20r$, $a_{2}=2r,\,a_{3}=10r$
(diamond).


\newpage

\begin{references}
\bibitem{ENCYC}  {\em The Encyclopedia of Molecular Biology}, Ed. J. Kendrew
(Blackwell, Oxford, 1994).

\bibitem{White}  H. Qian and J.H. White, {\em J. Biomol. Struct. and Dyn.} 
{\bf 16}, 663 (1998).

\bibitem{Zajac}  E.E. Zajac, {\em Transactions of the ASME} p. 136 (March
1962).

\bibitem{PG}  P.-G. deGennes, {\em Scaling Methods in Polymer Physics}
(Cornell Univeristy Press, Ithaca, 1979).

\bibitem{Volog}  A. A. Podtelezhnikov and A.V. Vologodskii, {\em %
Macromolecules} {\bf 33}, 2767 (2000).

\bibitem{Albert}  N.L. Goddard, G. Bonnet, O. Krichevsky, and A. Libchaber, 
{\em Phys. Rev. Lett.} {\bf 85}, 2400 (2000).

\bibitem{PRL2000}  S. Panyukov and Y. Rabin {\em Phys. Rev. Lett.} {\bf 85},
2404 (2000); {\em Phys. Rev. E }{\bf 62}, 7135 (2000).

\bibitem{PRL2001}  Y. Rabin and S. Panyukov, submitted to {\em Phys. Rev.
Lett.}

\bibitem{Love}  A.E.H. Love, {\em A Treatise on the Mathematical Theory of
Elasticity} (Dover, New York, 1944).

\bibitem{Shape}  J.J. Koenderink, {\em Solid Shape} (MIT Press, Cambridge,
1990).

\bibitem{Chaikin}  P.M Chaikin and T.M. Lubensky, {\em Principles of
Condensed Matter Physics} (Cambridge University Press, Cambridge, 1985),
Chapter 6.

\bibitem{Tobias1}  I. Tobias, B.C. Coleman, and M. Lembo, {\em J. Chem. Phys.%
} {\bf 105}, 2517 (1996).

\bibitem{Tobias2}  B.C. Coleman, M. Lembo, and I. Tobias, {\em Meccanica} 
{\bf 31}, 565 (1996).

\bibitem{Mezard}  C. Bouchiat and M. Mezard, {\em Phys. Rev. Lett.} {\bf 80}%
, 1556 (1998).

\bibitem{Fuller}  F.B. Fuller, {\em Proc. Natl. Acad. Sci.} {\bf 68}, 815
(1971); {\em Proc. Natl. Acad. Sci.} {\bf 75}, 3557 (1975).

\bibitem{Tobias}  We would like to thank I. Tobias and A. Maggs for bringing
this fact to our attention.

\bibitem{Topology}  J.H. White {\em Am. J. Math.} {\bf 91,} 693 (1969).

\bibitem{Maggs}  A.C. Maggs, {\em Phys. Rev. Lett.} {\bf 85}, 5472 (2000);
cond--mat/0009182.

\bibitem{Beard}  D.A. Beard and T. Schlick, {\em J. Chem. Phys.} {\bf 112},
7323 (2000).

\bibitem{Doi}  M. Doi and S.F. Edwards, {\em The Theory of Polymer Dynamics }%
(Oxford University Press, Oxford, 1986).

\bibitem{GoGo-PRL-98}  R.E. Goldstein, T.R. Powers, and C.H. Wiggins, {\em %
Phys. Rev. Lett.} {\bf 80}, 5232 (1998).

\bibitem{MoNe-Ma-98}  J. D. Moroz, P. Nelson, {\em Macromolecules} {\bf 31},
6333, (1998).

\bibitem{Kremer}  T.B. Liverpool, R. Golestanian, and K. Kremer, {\em Phys.
Rev. Lett. }{\bf 80}, 405 (1998).

\bibitem{Haijun}  Z. Haijun, Z. Yang, and O.-Y. Zhong-can, {\em Phys. Rev.
Lett.} {\bf 82}, 4560 (1999).

\bibitem{Liverpool}  R. Golestanian and T.B. Liverpool, {\em Phys. Rev. E }%
{\bf 62}, 5488 (2000).

\bibitem{miniplasmids}  I. Tobias, {\em Biophys. J.} {\bf 74}, 2545 (1998); 
{\em J. Chem. Phys.} {\bf 113}, 6950 (2000).
\end{references}
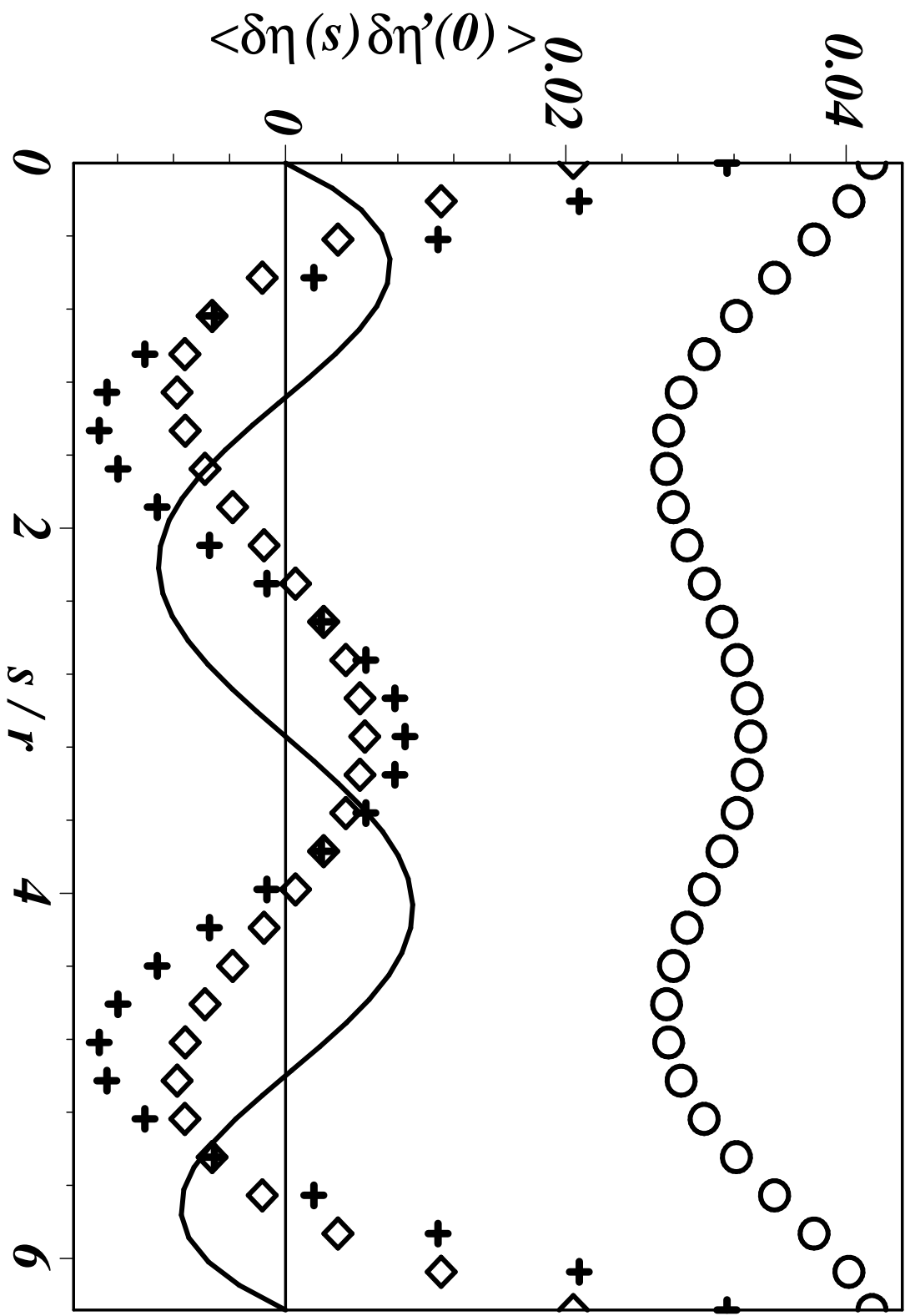
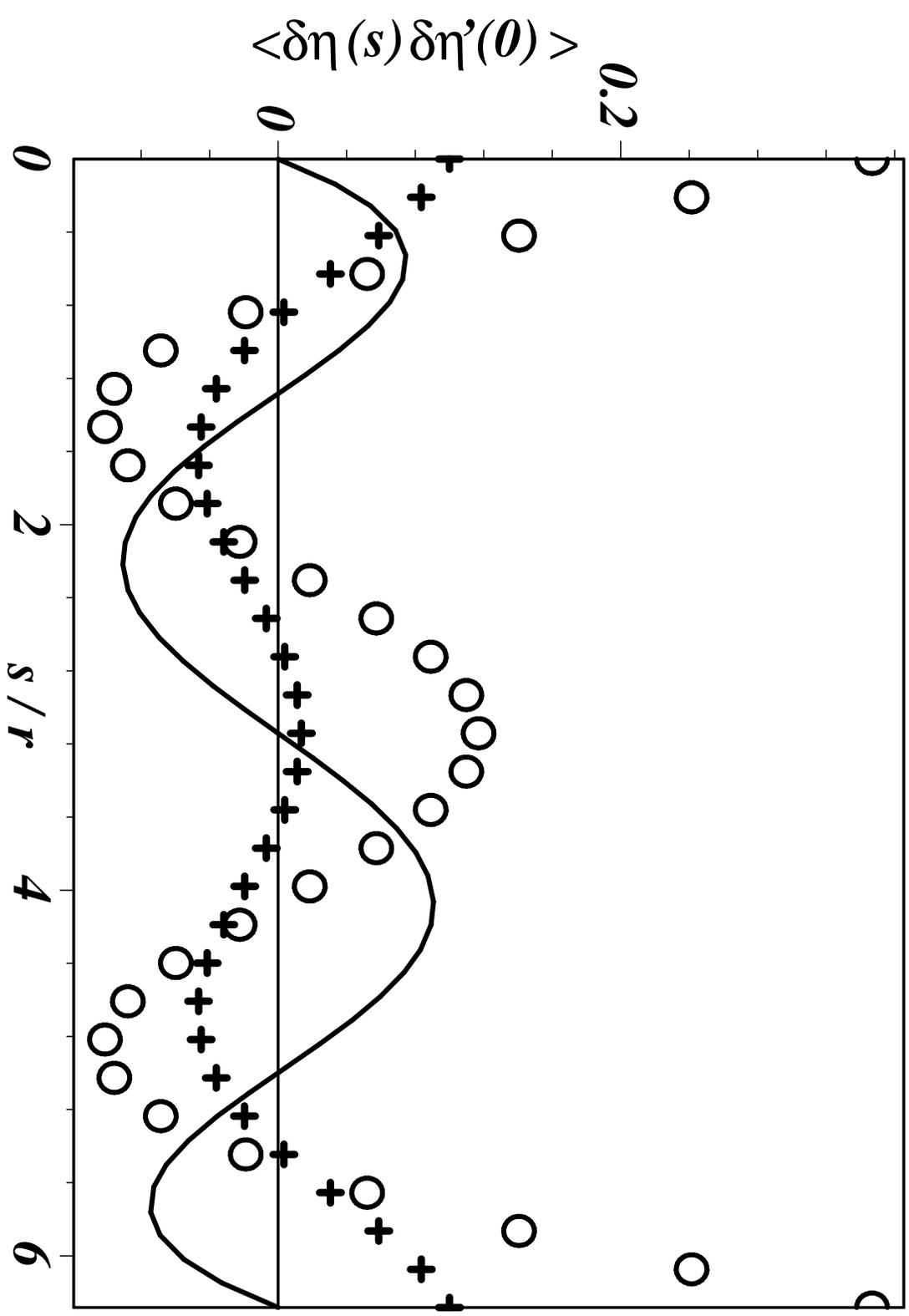
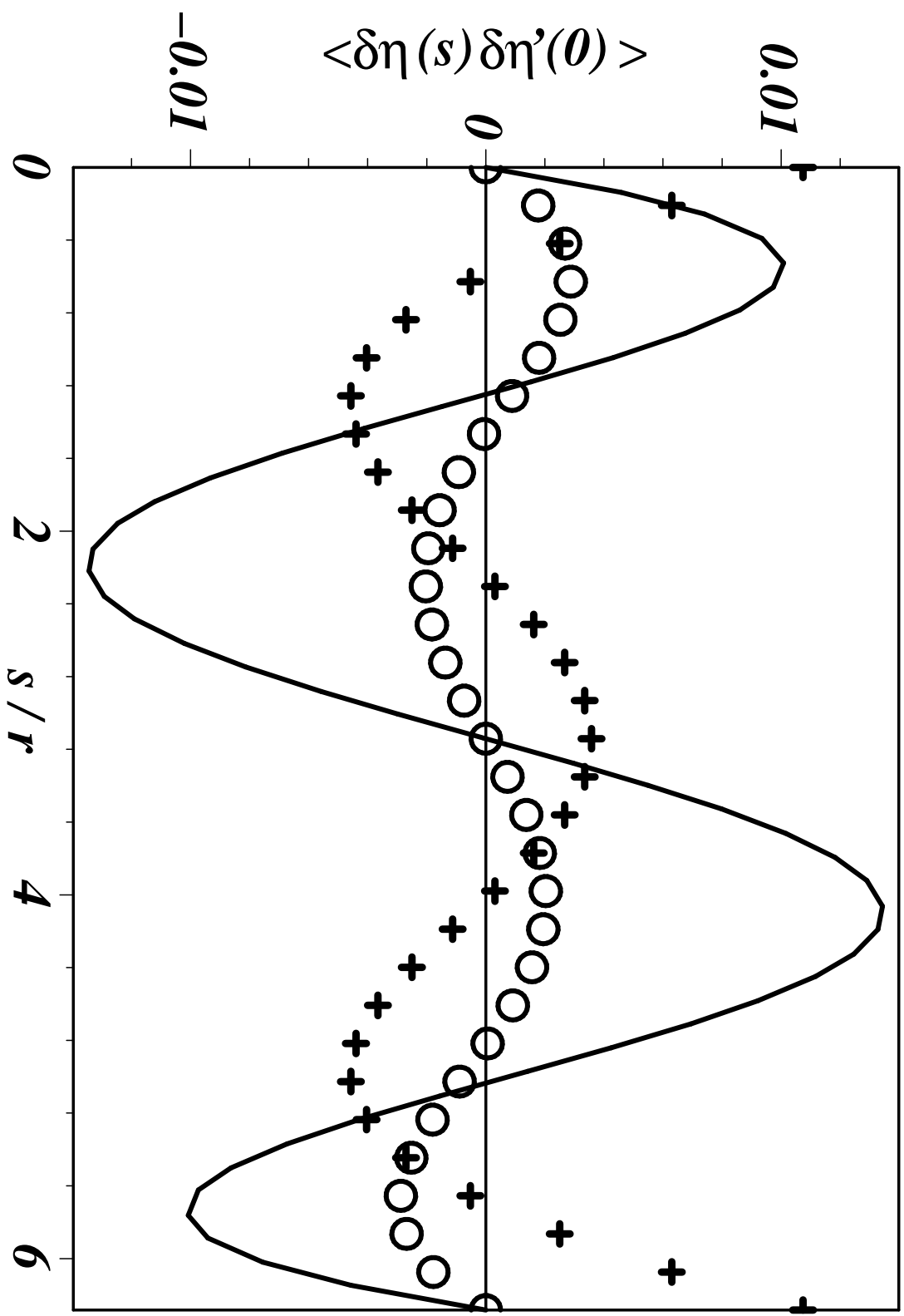
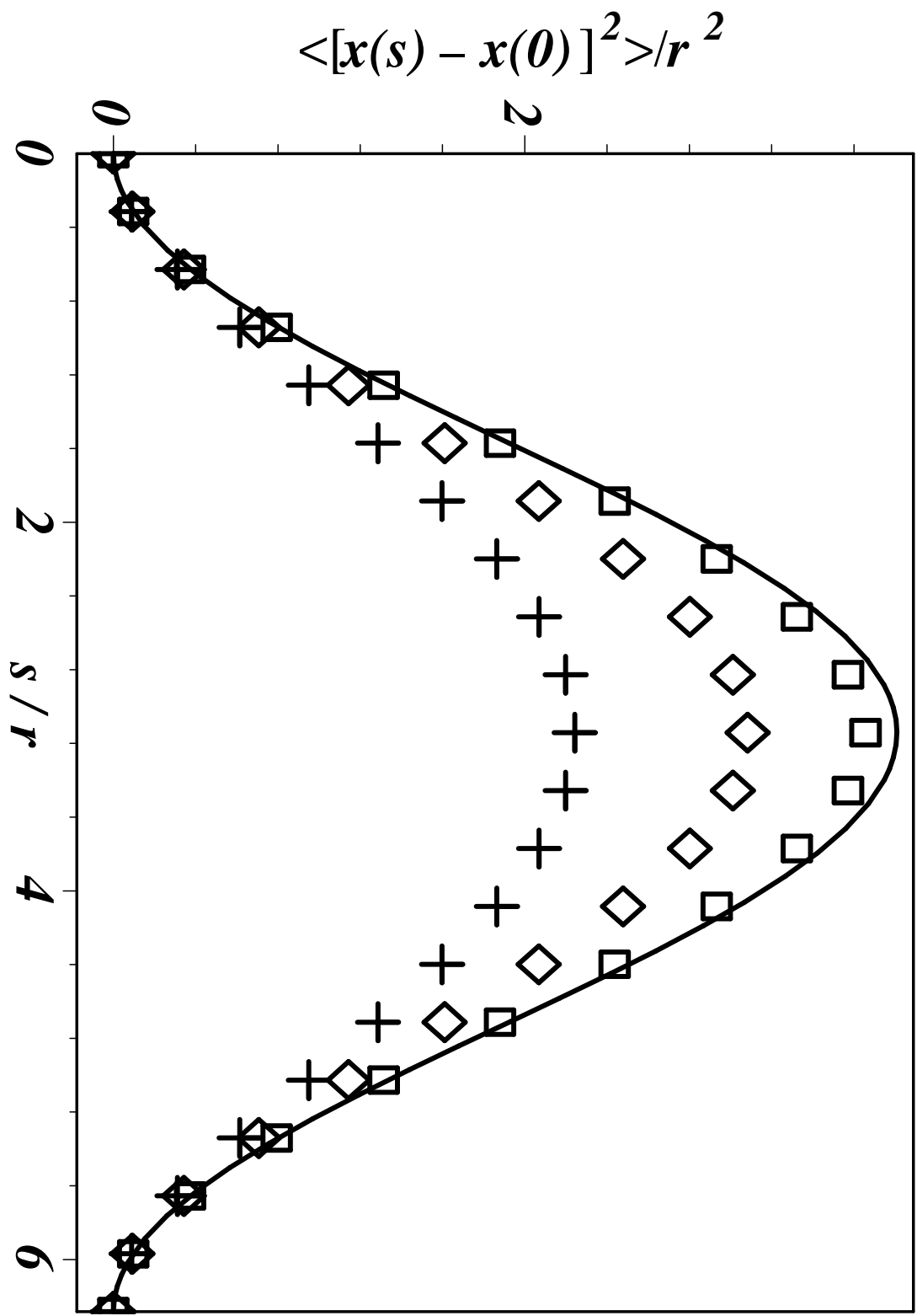
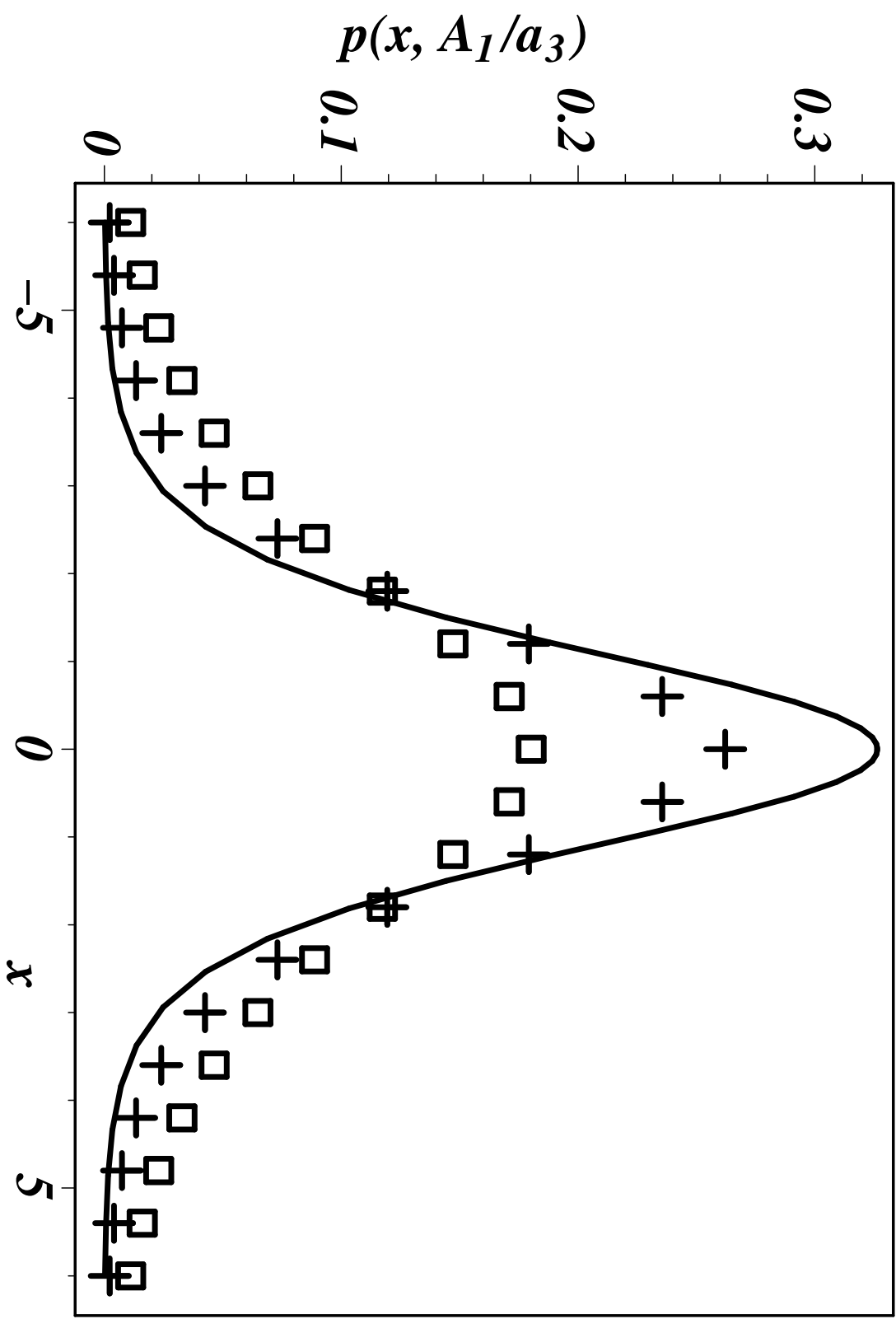
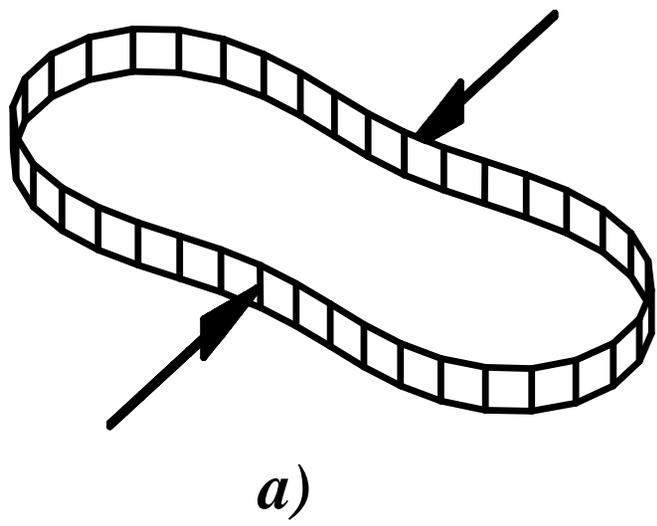
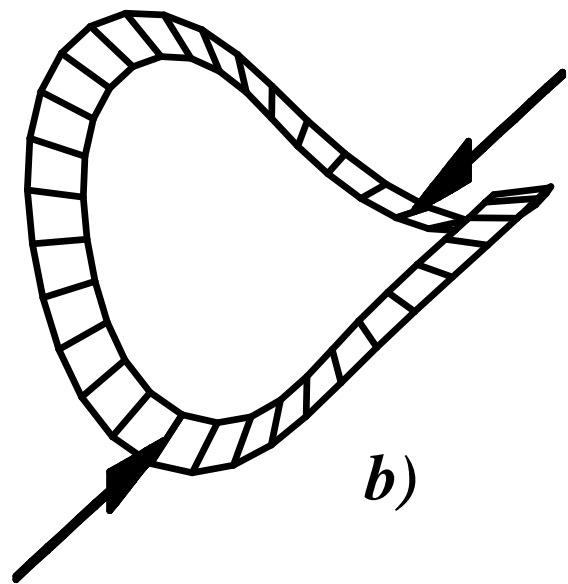
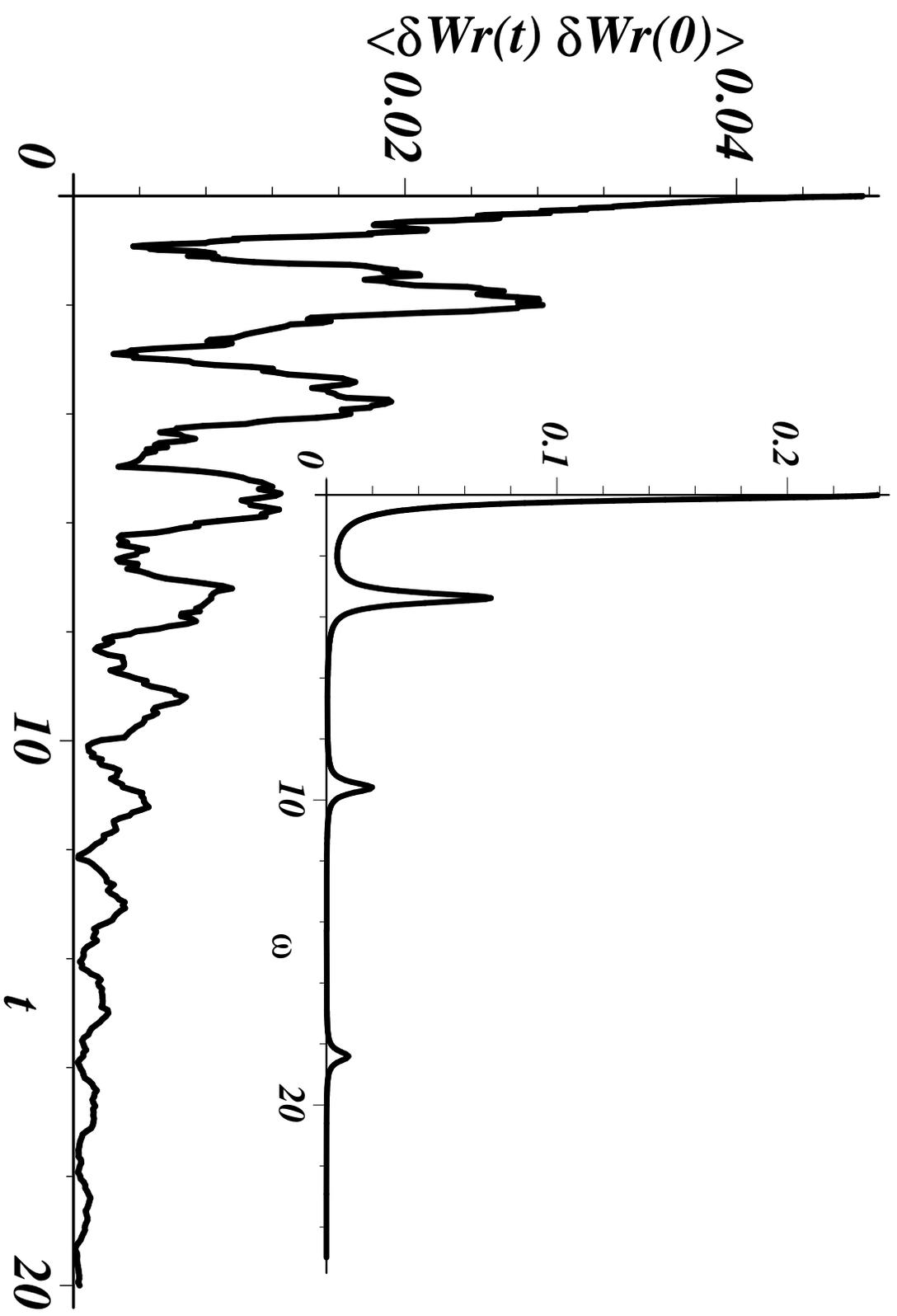
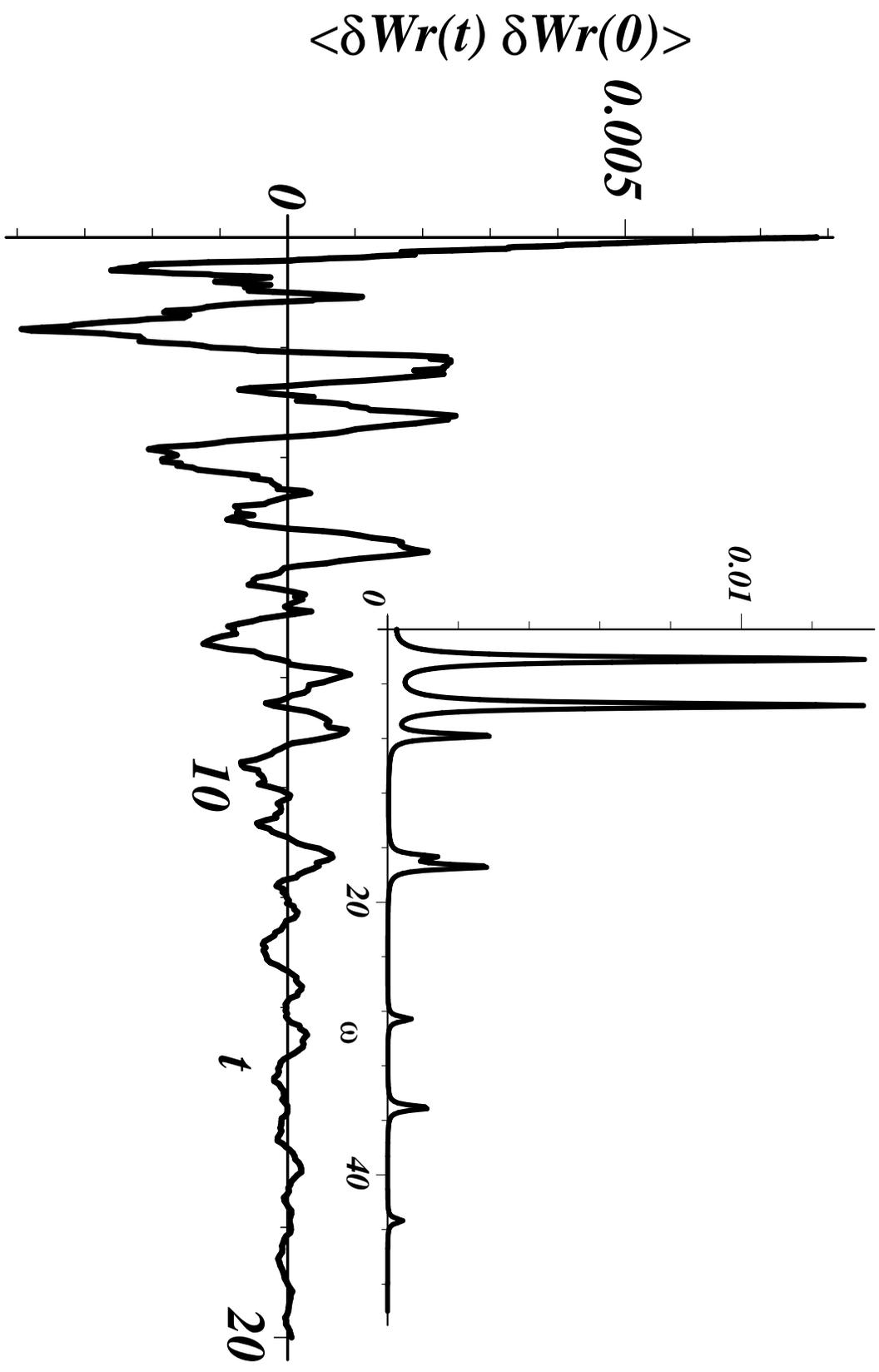
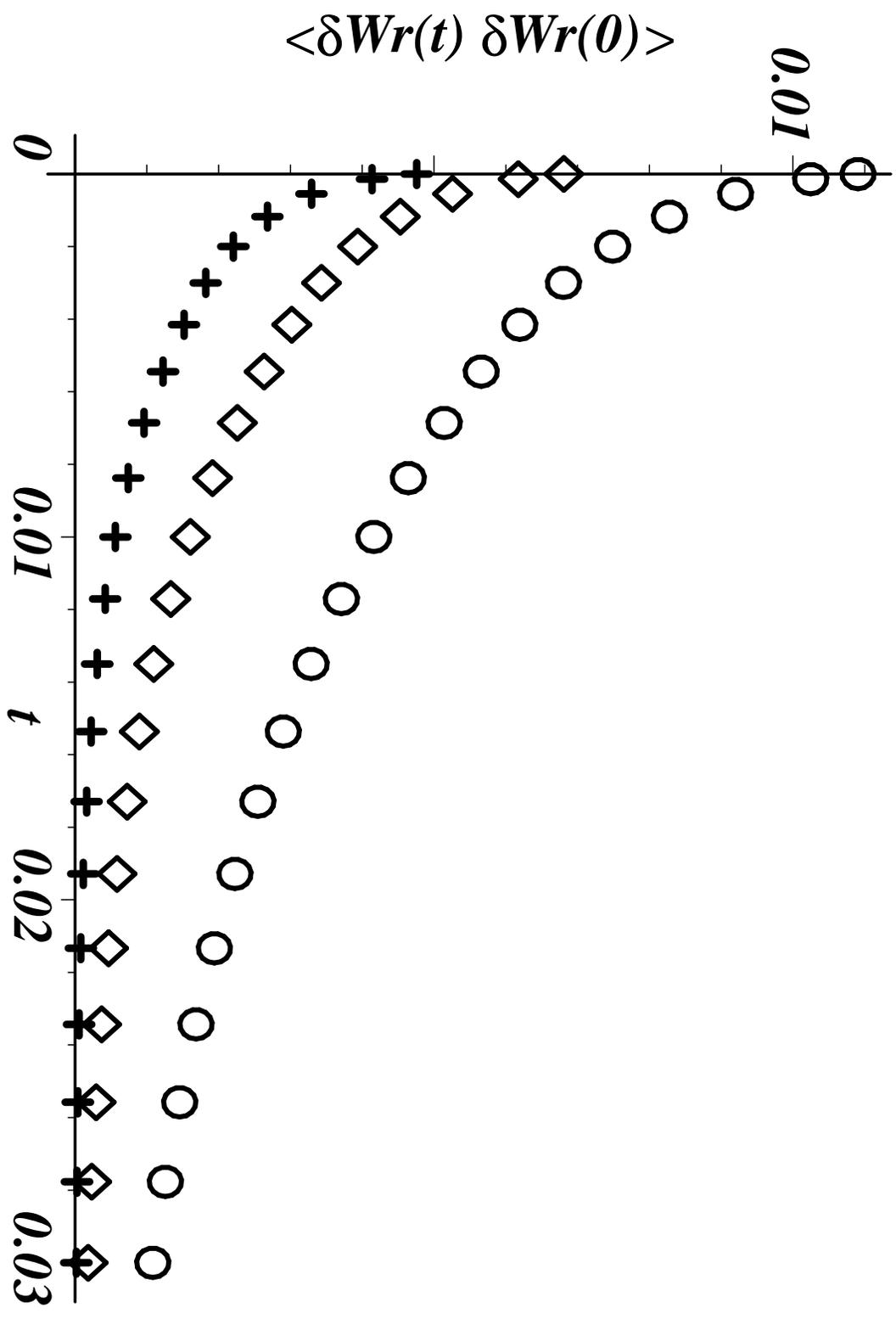
\end{document}